\documentclass[a4paper,twocolumn,11pt,accepted=2024-09-09]{quantumarticle}
\pdfoutput=1
\usepackage[utf8]{inputenc}
\usepackage[english]{babel}
\usepackage[T1]{fontenc}
\usepackage{amsmath}
\usepackage{amssymb}
\usepackage{hyperref}

\usepackage{natbib}

\usepackage{lipsum}

\usepackage[caption=false]{subfig}
\usepackage[ruled,linesnumbered]{algorithm2e}

\usepackage{booktabs}   % Improves table formatting
\usepackage{tabularx}   % Allows for adjustable column size within tables
\usepackage{multirow}   % Allows for rows spanning multiple rows in tables
\usepackage[para,flushleft,online]{threeparttable}  % Allows for nice tables
\usepackage{makecell}   % Multiline table cells
\usepackage{siunitx}    % SI unit support

\usepackage{graphicx} % include for figures
\graphicspath{{./figures/}} % set path for all graphics

\usepackage{xcolor}
\usepackage{ulem}  % Strikethrough text
\usepackage{cleveref}
\usepackage{tikz}
\usetikzlibrary{quantikz}

\renewcommand{\arraystretch}{1.5}
\usepackage{physics,braket,bm} % physics and math
\usepackage{mathtools} % for the \prescript command

\renewcommand{\o}{\text{o}}

\newcommand{\mcS}{\mathcal{S}}

\newcommand{\mcF}{\mathcal{F}}

\newcommand{\ca}{{\sim}}
\newcommand{\mc}[1]{\multicolumn{1}{c}{#1}}
\newcommand{\ms}[1]{\midstick[#1 ,brackets=none]{=}\qw}
\newcommand{\mcnot}[1]{\textsc{c}^{#1}\textsc{not}}

\DeclarePairedDelimiter\ceil{\lceil}{\rceil}
\DeclarePairedDelimiter\floor{\lfloor}{\rfloor}

\usepackage[inline]{enumitem} % in-line lists and \setlist{} (below)

\begin{document}

\title{Quantum-classical tradeoffs and multi-controlled quantum gate decompositions in variational algorithms }

%\author{Lídia del Rio}
%\affiliation{Institute for Theoretical Physics, ETH Zurich, 8093 Zurich, Switzerland}
%\orcid{0000-0002-2445-2701}
%\author{Christian Gogolin}
%\email{latex@quantum-journal.org}
%\homepage{http://quantum-journal.org}
%\orcid{0000-0003-0290-4698}
%\thanks{You can use the \texttt{\textbackslash{}email}, \texttt{\textbackslash{}homepage}, and \texttt{\textbackslash{}thanks} commands to add additional information for the preceding \texttt{\textbackslash{}author}. If applicable, this can also be used to indicate that a work has previously been published in conference proceedings.}
%\affiliation{Covestro Deutschland AG, Kaiser-Wilhelm-Allee 60, 51373 Leverkusen, Germany}
%\author{Marcus Huber}
%\affiliation{Institute for Quantum Optics \& Quantum Information (IQOQI), Austrian Academy of Sciences, Boltzmanngasse 3, Vienna A-1090, Austria}
%\orcid{0000-0003-1985-4623}
%\author{Cassandra Granade}
%\affiliation{Microsoft Research, Quantum Architectures and Computation Group, Redmond, WA 98052, USA}
%\author{Johannes Jakob Meyer}
%\affiliation{Dahlem Center for Complex Quantum Systems, Freie Universität Berlin, 14195 Berlin, Germany}
%\orcid{0000-0003-1533-8015}
%\author{Victor V. Albert}
%\affiliation{Institute for Quantum Information and Matter \& Walter Burke Institute for Theoretical Physics, Caltech, Pasadena, CA 91125, USA}
%\orcid{0000-0002-0335-9508}

\author{Teague Tomesh}
\orcid{0000-0003-2610-8661}
\email{teague.tomesh@infleqtion.com}
\affiliation{Princeton University, Princeton, NJ, 08540, USA}
\affiliation{Infleqtion, Chicago, IL, 60604, USA}

\author{Nicholas Allen}
\orcid{0000-0002-7651-0811}
%\email{nicholas.allen@uwaterloo.ca}
\affiliation{Princeton University, Princeton, NJ, 08540, USA}
\affiliation{University of Waterloo, Institute for Quantum Computing, Waterloo, ON N2L 3G1, Canada}

\author{Daniel Dilley}
\orcid{0000-0002-8821-4059}
%\email{ddilley@anl.gov}
\affiliation{Quantum Quadrivium Technologies LLP, Evergreen Park , IL, 60805, USA}
\affiliation{Argonne National Laboratory, 9700 S.~Cass Avenue, Lemont, IL, 60439, USA}

\author{Zain H.~Saleem}
\orcid{0000-0002-8182-2764}
\email{zsaleem@anl.gov}
\affiliation{Argonne National Laboratory, 9700 S.~Cass Avenue, Lemont, IL, 60439, USA}

\maketitle

\begin{abstract}
The computational capabilities of near-term quantum computers are limited by the noisy execution of gate operations and a limited number of physical qubits. Hybrid variational algorithms are well-suited to near-term quantum devices because they allow for a wide range of tradeoffs between the amount of quantum and classical resources used to solve a problem. 
This paper investigates tradeoffs available at both the algorithmic and hardware levels by studying a specific case --- applying the Quantum Approximate Optimization Algorithm (QAOA) to instances of the Maximum Independent Set (MIS) problem.
We consider three variants of the QAOA which offer different tradeoffs at the algorithmic level in terms of their required number of classical parameters, quantum gates, and iterations of classical optimization needed.
Since MIS is a constrained combinatorial optimization problem, the QAOA must respect the problem constraints. This can be accomplished by using many multi-controlled gate operations which must be decomposed into gates executable by the target hardware.
We study the tradeoffs available at this hardware level, combining the gate fidelities and decomposition efficiencies of different native gate sets into a single metric called the \textit{gate decomposition cost}. %With this metric, we find that native entangling gates with higher fidelity, but low degree, can outperform lower fidelity entangling gates with higher degree.
%With this metric, we find that native entangling gates with lower fidelity, but high degree, can outperform higher fidelity entangling gates with lower degree because of the efficient gate decompositions they enable.
\end{abstract}
    
\maketitle
    
    %%%%%%%%%%%%%%%%%%%%%%%%%%%%%%%%%%%%%%%%%%%%%%%%%%%%%%%%%%%%
\section{Introduction}
    
Today's quantum computing platforms are capable of executing single- and two-qubit gates with fidelities that have steadily improved year over year \cite{gambetta2017building, wright2019benchmarking, arute2019quantum, arrazola2021quantum}.
However, implementing multi-qubit operations between three or more qubits presents significant challenges, although some progress has been made in this direction \cite{monz2009realization, isenhower:2011}, and it remains unclear what level of fidelity must be achieved before multi-qubit operations may outperform one- and two-qubit operations when executing a quantum algorithm. 
%we still have some time before quantum hardware is capable of natively and accurately implementing large multi-qubit gates.
%It is therefore important to study the different decompositions of these large multi-qubit gates into single-qubit, two-qubit and smaller multi-qubit gates that are natively implementable on current and near-term quantum devices.

The limitations of today's Noisy Intermediate-Scale Quantum (NISQ) computers severely restrict the set of executable quantum algorithms \cite{preskill2018quantum}. 
Hybrid variational algorithms are a class of applications which are well-suited to NISQ systems because they make use of both quantum and classical resources when solving a problem, and this allows the quantum computer to focus on a single, well-defined task \cite{cerezo2021variational}.
For a given computational workload, a variational algorithm may make algorithmic level tradeoffs (e.g., varying the number of classical parameters, the size of the variational ansatz, or iterations of classical optimization) that exchange quantum and classical resources.
Furthermore, current NISQ computers support a wide variety of native gate sets which enables different choices for decomposing the algorithmic level circuits into hardware executables \cite{murali2019full}.

The Quantum Approximate Optimization Algorithm (QAOA) \cite{farhi2014quantum} is a hybrid variational algorithm that solves both constrained and unconstrained combinatorial optimization problems. 
%The application of the QAOA to unconstrained optimization problems such as MaxCut requires only single- and two-qubit gates.
For constrained combinatorial optimization problems there are two methods for dealing with the added problem constraints. One choice is to convert the constrained problem into an unconstrained one by introducing a Lagrange multiplier into the objective function. This can then be executed on quantum hardware using only single- and two-qubit gates, however, this method often requires an additional hyper-parameter optimization step in order to correctly set the value of the Lagrange multiplier \cite{saleem2020approaches}. Alternatively, the constraint may be introduced within the ansatz itself which then requires the use of multi-controlled quantum gates \cite{hadfield2018quantum,hadfield2019quantum,saleem2020max}. The constraint-aware ansatz ensures that the constraints are satisfied at all times during the optimization and the extremization is only performed over the set of feasible states. In this paper we focus solely on the latter case which uses the constraint-aware ansatz and therefore requires the use of multi-controlled gates.

%\textcolor{orange}{1st paragraph: Different ways to decompose multi-qubit gates and the final cost will depend on the native gate set of the hardware. Transition sentence to next paragraph: "you can choose to stop the decomposition early if your hardware supports it". All of the decompositions we study are linear but the constant scaling factors vary greatly depending on the native gate set of the hardware.}
%In addition to the tradeoffs that can be made at the algorithmic level of the QAOA, choices made at the compiler and hardware levels can greatly impact the overhead gate cost when decomposing the quantum circuits into hardware-executable gate sets.
Before the QAOA may be executed, the algorithmically defined multi-qubit gates must be decomposed into equivalent sets of gate operations which are natively supported by the quantum hardware. Depending on the specific native gate set, the final overhead gate cost incurred by the decomposition may vary greatly.
Quantum gate decompositions are a broad and extensive area of study \cite{barenco1995elementary, vartiainen2004efficient, martinez2016compiling, gokhale2019asymptotic, rakyta2022approaching}. Strategies for general unitary decompositions include algorithmic rule-based constructions \cite{barenco1995elementary, vartiainen2004efficient, li2013decomposition}, direct synthesis via numerical optimization \cite{younis2021qfast}, and optimal control \cite{shi2019optimized}. 
Oftentimes the multi-qubit gates appearing in the constrained QAOA contain a large amount of structure (e.g., they may be expressed simply as sets of single-qubit rotations and multi-controlled  \textsc{not}'s) that allows for more efficient decompositions \cite{barenco1995elementary, gokhale2019asymptotic}.
%In the case of the constrained QAOA we are aided by the fact that the basic building blocks of the ansatzes, the partial mixers, appear as multi-qubit gates with many controls acting on a single qubit.
%The large amount of structure in these unitary gates allows for more efficient decompositions into basic single-, two-, and three-qubit gates.
%For example, because the partial mixers, which appear as multi-controlled single-qubit rotations, can be decomposed into groups of single-qubit gates and multi-controlled \textsc{not}'s, we can rely on the vast amount of prior work which specifically study efficient decompositions of these gates \cite{barenco1995elementary, gokhale2019asymptotic}.
%However, in practice, quantum compilers must account for one additional constraint when solving the decomposition problem: all of the basis gates must be operations that are natively supported by the hardware. The final overhead gate cost is strongly dependent on the gate set offered by the underlying hardware. 

%%%%%%%%%%%%%%%%%%%%%%%%%%%
In this paper, we study the quantum and classical tradeoffs available to variational quantum algorithms, specifically focusing on the case of constrained QAOA applied to Maximum Independent Set (MIS) problems. We consider both algorithmic and hardware level design choices that influence the final amount of classical and quantum resources that are required.   

At the algorithmic level, we consider three variants of the QAOA which differ in the number of classical parameters, rounds of optimization, and gate operations they require.
In the original formulation of the QAOA, the ansatz is composed of alternating layers of parameterized gates and all of the gates within a single layer are parameterized by the same variable. We refer to this version as single-angle QAOA (SA-QAOA).
Recent work has shown that, for both unconstrained optimization problems, such as MaxCut, and constrained problems, such as MIS, parameterizing the gates within each layer by a vector of variables can result in higher approximation ratios and shorter circuits at the expense of additional classical parameters~\cite{herrman2022multi, saleem2020approaches}. We refer to this implementation of the QAOA as multi-angle QAOA (MA-QAOA).
Finally, the number of multi-controlled quantum gates in both the SA-QAOA and MA-QAOA scales linearly with the problem size. To mitigate this large quantum resource requirement, a new approach called the Dynamic Quantum Variational Ansatz (DQVA) was introduced in \cite{saleem2020approaches}. The DQVA removes this scaling dependency on the problem size by making the number of parameters and multi-controlled gates a tunable hyper-parameter in exchange for additional iterations of classical optimization. 

At the hardware level, we study the cost of decomposing the large multi-qubit gates into different native gate sets containing different levels of controlled gates and higher dimensional qudits.
Our consideration of native gate sets with higher degree multi-qubit gates is motivated by the recent emergence of quantum hardware capable of supporting such operations. These include neutral atoms \cite{isenhower:2011, saffman2016quantum}, trapped ions \cite{espinoza:2021, katz:2022, rasmussen:2020}, and recent demonstrations of Toffoli gates in superconducting quantum hardware \cite{kim2022high}. Simultaneously, multiple experiments have demonstrated the viability of using \textit{qutrits}, containing three logical states $\ket{0}$, $\ket{1}$, and $\ket{2}$, within quantum computations~\cite{morvan2021qutrit, gokhale2020optimized}.
Prior work has already shown that more expressive gate sets, containing multiple two-qubit entangling operations from different gate families, are beneficial for the compilation of quantum applications \cite{lao2021designing}. We extend this intuition here by studying the impact that natively supported multi-controlled gates and three-level qutrits can have on the decomposition overheads for the QAOA.

In summary, our contributions include:

\begin{itemize}
    \item Numerical simulations of the QAOA on 20-node graphs, highlighting the tradeoffs that can be made between classical and quantum resources while achieving similar performance amongst the three QAOA variants.
    \item Asymptotic gate counts with exact scaling coefficients for the decomposition of the multi-controlled single-qubit rotations appearing in the constrained QAOA circuits. We consider scenarios where varying numbers of ancillae and different native gate sets are made available for the decompositions.
    \item We introduce a new metric, the gate decomposition cost, which shows that we can use lower fidelity native gates in one QAOA to outperform the higher fidelity gates in another.
    %, as long as the gate cost is below a certain threshold native entangling gates with higher fidelity, but low degree, can outperform lower fidelity entangling gates with higher degree. which shows that natively supported multi-qubit gates --- even if their fidelity is lower than the native two-qubit gates --- can produce overall more efficient quantum circuits.
\end{itemize}

The goal of this paper is to aid future system designs by calculating the fidelity of native multi-qubit (or qutrit) gates that is necessary to achieve performance improvements over the typical two-qubit gates for realistic workloads.

The rest of the paper is organized as follows, we begin in Section \ref{QAOA} by introducing the different flavors of the constrained QAOA applied to the MIS problem. Section \ref{decompositions} describes the different decompositions of multi-controlled single-qubit rotations. In Section \ref{resource} we compare the variants of the constrained QAOA and perform resource analysis. Finally, in Section \ref{conclusions} we summarize our conclusions and suggest future directions.

\section{Quantum Approximate Optimization for Maximum Independent Set}\label{QAOA}
\begin{table}[t]
\tabcolsep=0.24cm
\begin{threeparttable}
    \begin{tabular}{c c c}
    \toprule
    \textbf{Algorithm} & \textbf{\begin{tabular}[c]{@{}c@{}}Number of \\ parameters, \\ $\abs{\boldsymbol{\theta}}$\end{tabular}} & \textbf{\begin{tabular}[c]{@{}c@{}}Rounds of \\ variational \\ optimization\end{tabular}} \\ 
    \midrule
    SA-QAOA            & \begin{tabular}[c]{@{}c@{}}Constant,\\ $\abs{\boldsymbol{\theta}}=2p$\end{tabular}                                                                    & Single                                                                                    \\
    MA-QAOA            & \begin{tabular}[c]{@{}c@{}}Scales with\\ problem size,\\ $\abs{\boldsymbol{\theta}} = \text{up to } 2pn $\end{tabular}       & Single                                                                                     \\ 
    DQVA               & \begin{tabular}[c]{@{}c@{}}Tunable with\\ problem size,\\ $\abs{\boldsymbol{\theta}} = \nu$\end{tabular}       & Multiple                                                                                  \\
    \bottomrule
    \end{tabular}
\end{threeparttable}
\caption{The three flavors of the QAOA considered in this work. Each makes different tradeoffs with respect to the classical (e.g., number of parameters, rounds of optimization) and quantum (e.g., circuit depth) resources they require. We consider the Maximum Independent Set (MIS) problem over graphs with $n$ nodes.}
\label{tab:overview}
\end{table}

The QAOA is a hybrid quantum-classical algorithm developed for the purpose of finding approximate solutions to combinatorial optimization problems. The objective of the QAOA is to prepare the ground state of a problem-dependent quantum operator using a variational ansatz, i.e., a parameterized quantum circuit. This quantum operator is defined on $n$-bit strings and is diagonal in the computational basis
\begin{equation}\label{qoperator}
    C_{obj} | b \rangle = C(\textbf{b}) |b \rangle
\end{equation}
where $\textbf{b} = \{b_1,b_2,b_3 \dots b_n\} \in \{0,1\}^n$. The expectation value of this operator is measured on a quantum computer after preparing the state 
\begin{equation}
    |\psi_p(\boldsymbol{\gamma},\boldsymbol{\beta} )\rangle = e^{-i\beta_p
        M}e^{-i\gamma_p C}\cdots e^{-i\beta_1 M}e^{-i\gamma_1 C}|s\rangle
\end{equation}
where $|s\rangle$ is the initial state, $e^{i \gamma C}$ is the phase separator unitary, diagonal in computational basis, $e^{i \beta M}$ is the mixing unitary that mixes the state among one another during optimization and $p$ controls the number of times the unitary operators are applied. The expectation value of $C_{obj}$ with respect to this variational state,
\begin{equation}
E_p(\boldsymbol{\gamma},\boldsymbol{\beta})=\langle  \psi_p(\boldsymbol{\gamma},\boldsymbol{\beta} )|C_{obj}|\psi_p(\boldsymbol{\gamma},\boldsymbol{\beta} )\rangle , 
\end{equation}
is evaluated on a quantum computer before being passed to a classical optimizer which attempts to find the optimal parameters which maximize $E_p(\boldsymbol\gamma, \boldsymbol\beta)$.
This maximization is achieved for the states corresponding to the optimal solutions of the original optimization problem.
    
\subsection{Unconstrained QAOA}
The Maximum Independent Set (MIS) problem involves finding the largest set of nodes in a graph such that no two nodes are connected to each other. 
Only a subset of all possible solutions (i.e., all $n$-bit strings) are feasible solutions which satisfy the problem constraints. However, we may still apply the QAOA to this problem by adding a penalty term to the objective function
\begin{equation}
    C_{obj}=H - \lambda \; C_{pen} = \sum_{i\in V} b_i - \lambda \sum_{ i,j \in E} b_i b_j ,
\end{equation}
and
\begin{equation}
    b_i= \frac{1}{2} \left(1- Z_i\right)
\end{equation}
where $V$ and $E$ are the set of all the nodes and edges in the graph, $Z_i$ is the Pauli-Z operator acting on the $i$th qubit and $\lambda$ is the Lagrange multiplier.
The first term $H$ is the Hamming weight of the bit string and the second term $C_{pen}$ is the penalty term which penalizes every time two nodes are connected by an edge.
This objective function defines an unconstrained optimization problem which can be solved by the QAOA and can be implemented using only single- and two-qubit gates, but it has the disadvantage that the optimization is performed over both feasible and infeasible states. Finding the optimal value for $\lambda$, which effectively balances the trade-off between optimality and feasibility in the objective function, can pose a considerable challenge. %In this paper we will focus on variants of the QAOA which enforce the problem constraints at the circuit level using multi-controlled gates to ensure that optimization remains within the space of feasible states. 

\subsection{Constrained QAOA}\label{QAO-Ansatz}
\begin{figure}[t]
    \centering
    \resizebox{\linewidth}{!}{%
    \begin{quantikz}[column sep=0.2cm]
        & \octrl{1} & [0.1em] \ms{4} & \qw & \gate{X} & \ctrl{3} & \qw & \ctrl{3} & \gate{X} & \qw \\
        & \octrl{1} & \qw & \qw & \gate{X} & \ctrl{2} & \qw & \ctrl{2} & \gate{X} & \qw \\
        & \octrl{1} & \qw & \qw & \gate{X} & \ctrl{1} & \qw & \ctrl{1} & \gate{X} & \qw \\
        & \gate{R_x\qty(\theta)} & \qw & [0.1em] \gate{R_z\qty(\frac{\pi}{2})} & \gate{R_y\qty(\frac{\theta}{2})} & \targ{} & \gate{R_y\qty(\frac{-\theta}{2})} & \targ{} & \gate{R_z\qty(\frac{-\pi}{2})} & \qw
    \end{quantikz}
    }
    \caption{Decomposition of the partial mixers used in the constrained QAOA for MIS into two multi-controlled Toffolis and single-qubit rotations \cite{barenco1995elementary}. The partial mixer enforces the MIS constraint by only performing the single-qubit rotation if all neighbors are in the $\ket{0}$ state.}
    \label{fig:partialmixer}
\end{figure}
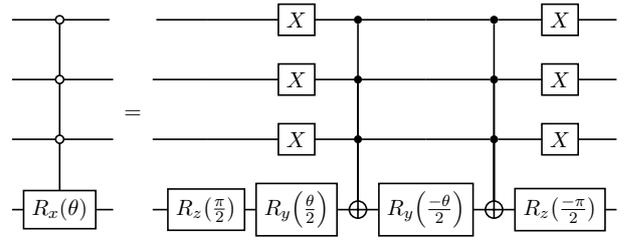

In constrained versions of the QAOA the structure of the ansatz is altered to ensure that the variational optimization takes place only over the set of feasible solutions \cite{hadfield2018quantum, hadfield2019quantum}, whereas the Lagrange multiplier method includes optimization over a larger set of states that may not correspond to an independent set. Since the problem constraints are handled at the circuit level, the objective function is simply the Hamming weight operator,
\begin{equation}
    C_{obj}= H = \sum_{i \in V} b_i .
    \label{qaoa objective}
\end{equation}
The initial state may be any feasible state or a superposition of feasible states. The phase separator unitary $U_{C}(\gamma) \coloneqq e^{i \gamma H}$ is determined by the objective function. The mixing unitary is specially designed to ensure that the ``independent set'' constraint is maintained at all times during optimization. Let $U_{M}(\beta) \coloneqq \prod_j e^{i \beta M_j}$,
where $M_j = X_j \bar{B} $ and we have defined
\begin{equation}
    \bar{B} \coloneqq  \prod_{k=1}^{\ell} \bar{b}_{v_k}, \;\;\;\;
    \bar{b}_{v_k}=\frac{1+Z_{v_k}}{2},
\end{equation}
where $v_k$ are the neighbors and $\ell$ is the number of neighbors for the
$k$th node. We can also write the mixer as
\begin{eqnarray}\label{eqn:sa-mixer}
\begin{aligned}
 U_M(\beta) & = \prod_{j=1}^{N}V_j(\beta) \\
            & = \prod_{j=1}^{N} \left(I + ( e^{-i\beta X_j}-I) \;  \bar{B} \right),
\end{aligned}
\end{eqnarray}
where we have used $\bar{b}_{v_j}^2=\bar{b}_{v_j}$.
The unitary mixer above is a product of $N$ partial mixers $V_i$, and in
general they may not all commute with each other $[V_i , V_j] \neq 0 $. 
The partial mixers are executed on a digital quantum computer using the circuit shown in Figure \ref{fig:partialmixer}.
    
%%%%%%%%%%%%%%%%%%%%%%%%%%%%%%%%%%%%%%%%%%%%%%%%%%%%%%%%%%%%

Additionally, the mixing unitary may be parameterized by a list of angles $\boldsymbol{\beta}$, such that each partial mixer can have a different classical variable as a parameter, and is defined up to a permutation of the partial mixers:
\begin{equation}\label{eqn:ma-mixer}
    U_M(\boldsymbol{\beta}) \simeq \mathcal{P}\big(V_1(\beta_1)V_2(\beta_2)\cdots
    V_N(\beta_N)\big).
\end{equation}
%For unconstrained optimization problems, such as MaxCut, prior work has found that allowing the mixing unitary to be parameterized by a vector of values (as in \eqref{eqn:ma-mixer}) can result in higher approximation ratios and shorter circuits at the expense of additional classical parameters~\cite{herrman2022multi}.
%In this work we apply these multi-angle mixers to the constrained optimization problem of MIS where, unlike the unconstrained mixer, the individual partial mixers do not commute with one another and we find performance which supports the results of \cite{herrman2022multi}.
To distinguish between the QAOA whose mixing unitary has a single angle, Eq.~\eqref{eqn:sa-mixer}, or multiple angles, Eq.~\eqref{eqn:ma-mixer}, we denote these cases as SA-QAOA and MA-QAOA, respectively.

\subsection{Dynamic Quantum Variational Ansatz}\label{DQVA}
An extension of the MA-QAOA, called the Dynamic Quantum Variational Ansatz (DQVA), was introduced in \cite{saleem2020approaches} which enables a tradeoff between the number of partial mixers and rounds of classical optimization.
The DQVA may be initialized in any superposition of feasible states, potentially even using the output of  a classical optimization algorithm $\ket{\textbf{c}} = \ket{c_{1} c_{2} \cdots c_{n}}$. This warm-start approach was recently proposed for Max-Cut and QUBOs in \cite{egger2021warm}.
Then, the phase separator and mixing unitaries are applied to construct the state,
\begin{eqnarray}\label{eqn:dqv-ansatz}
\begin{aligned}
    \psi(& \boldsymbol\gamma, \boldsymbol\alpha_1, \dots, \boldsymbol\alpha_p) = \\
    & U_C(\gamma_p)U_M^{\sigma_p}(\boldsymbol\alpha_p)\cdots U_C(\gamma_1)U_M^{\sigma_1}(\boldsymbol\alpha_1)\ket{\textbf{c}}
\end{aligned}
\end{eqnarray}
where, similarly to Eq.~\eqref{eqn:ma-mixer}, each of the individual mixers is given by
\begin{eqnarray}\label{eqn:permutation}
\begin{aligned}
    U_M^{\sigma_k}(& \boldsymbol\alpha_k) = \\
    & V_{\sigma_k(n)}\left(\alpha_k^{\sigma_k(n)}\right) \cdots
    V_{\sigma_k(1)}\left(\alpha_k^{\sigma_k(1)}\right)
\end{aligned}
\end{eqnarray}
for $k=1,2,\cdots,p$.
Here for simplicity we have fixed all permutations $\sigma_1=\sigma_2=\cdots=\sigma_p$, and henceforth drop the permutation index on mixers.

%Wherever the $i$-th bit of the initial state $c^i=1$, we set the parameter in the corresponding partial mixer to be $\alpha_k^i=0$, essentially ``turning off'' this partial mixer and setting it to the identity operator $I$.
%The key difference between Eq.~\eqref{eqn:permutation} and Eq.~\eqref{eqn:qao-mixer} is the mixer unitary is parameterized by a vector of angles instead of a single scalar.
In \cite{saleem2020approaches} it was found that the increased expressibility of Eq.~\eqref{eqn:permutation} resulted in improved performance with smaller circuit depth on constrained optimization problems, and in \cite{herrman2022multi} similar results were found for unconstrained problems. The increased expressibility of the unitary mixer has further advantages. We can set some of the partial mixers ``off'' by setting $\alpha_i = 0$, effectively reducing the quantum resources required by the ansatz. In this paper, we control the number of nonzero parameters in the variational ansatz, including both the phase separator $\gamma$ and mixer $\boldsymbol\beta$ parameters, using a hyperparameter $\nu$.

The final component of the DQVA algorithm is the dynamic ansatz update. This is a classical outer loop where in every iteration, called a \textit{mixer round}, the order of the partial mixers appearing in Eq.~\eqref{eqn:permutation} is randomly permuted. Each mixer round is composed of multiple \textit{inner rounds} where the ansatz is initialized with the best independent set found thus far, variationally optimized, and then uses the output of that optimization as the input for the next inner round. As more and more nodes become part of the independent set, the ansatz is continually updated as detailed in \cite{saleem2020approaches} until the entire graph has been traversed, increasing the size of the independent set at each iteration. 

The DQVA algorithm has previously been used in the quantum local search (QLS) approach in \cite{tomesh2022quantum}. In QLS the DQVA is applied to local neighbourhoods and the whole graph is traversed neighbourhood by neighbourhood, increasing the size of the independent set at each step. The DQVA was also applied in the quantum divide and conquer approach in \cite{saleem2021quantum} in which the graph is first partitioned and then the DQVA is implemented on each of the partitions separately. This allows the potential for distributed computing as well since each of the partitions can be run in parallel. 
The analysis of multi-controlled gate decompositions presented in this work is complimentary to these higher-level algorithmic approaches --- enabling more accurate resource estimates for these algorithms under different assumptions of the underlying native gate set.

\section{Multi-controlled Gate Decompositions}\label{decompositions}

Many quantum algorithms are expressed using multi-controlled quantum gates \cite{draper2004logarithmic, chakrabarti2008designing, arrazola2022universal, crow2016improved, premakumar2020measurement}. However, most currently available hardware platforms, like IBM or Rigetti's superconducting systems \cite{murali2019full} or IonQ's trapped ion systems \cite{ionq2022gates}, only expose single- and two-qubit primitive operations to the user. Thus, to execute large quantum gates on physical hardware, one needs to use a sequence of single- and two-qubit gates whose composite action is equivalent to the larger gate.

Methods for deriving these gate sequences are known as gate decomposition schemes. By imposing various restrictions on the physical system available to us (e.g. native gate set, number of available ancilla qubits), we can analyze the resources required to implement various operations by analyzing the gate complexity of these decompositions. We can then compare the resource efficiency between different algorithms or different physical systems, building upon the decomposition schemes given by Barenco \textit{et al.} in \cite{barenco1995elementary}.

\subsection{Definitions and notation}

In what follows, we denote a \(k\)-controlled \(U\) gate by \(C^k(U)\). We can then define the gate set \(\mcS^d_n\) to be the collection of quantum gates on $d$-level qudits given by
\begin{equation}
    \mcS^d_n = \qty{U(d), C(X), C^2(X), \ldots, C^{n-1}(X)},
\end{equation}
consisting of single qudit gates and every multi-controlled \textsc{not} from one control to \((n - 1)\) controls. 
The gates contained in any gate set are known as basis gates. We make the assumption that the available native gates on the hardware include only these generalized Toffoli gates, whereas, in reality, actual hardware may employ a variety of gates such as \textsc{cphase} gates, $\sqrt{\textsc{swap}}$ gates, cross-resonance gates, etc. Fortunately, this assumption does not usually pose a problem since any two-qubit entangling gate can be simulated by another two-qubit entangling gate with a constant overhead.
%We assume that these generalized Toffoli gates are the only two- and other multi-qubit gates natively available, which is not generally the case for real hardware, which may use \textsc{cphase} gates, $\sqrt{\textsc{swap}}$ gates, cross-resonance gates, etc. This is not generally an issue since any two-qubit entangling gate can be simulated by any other two-qubit entangling gate with constant overhead.
The primary gate sets we consider are:
\begin{enumerate}
    \item \(\mcS^2_2 = \qty{U(2), C(X)}\), the simplest universal gate set, and representative of what most architectures are capable of.
    \item \(\mcS^2_3 = \qty{U(2), C(X), C^2(X)}\), which includes native Toffoli gates.
    \item \(\mcS^3_2 = \qty{U(3), C(X_{+1}), C(X_{-1})}\), which consists of single- and two-\textit{qutrit} gates. See \cite{gokhale2019asymptotic} for further background information.
\end{enumerate}

When we consider the following gate decomposition schemes, we will generally assume that the ancilla qubits available for every computation are zeroed -- meaning that they start initialized to the \(\ket{0}\) state, and they are required to be returned to the \(\ket{0}\) state after each decomposition. This is to avoid the experimental overhead of needing to re-initialize the ancilla qubits back to \(\ket{0}\) in the middle of a computation. Here, we focus on the decomposition schemes for a \(C^n(U)\) gate with access to either one or \(n\) ancilla qubits.

\begin{figure}[t]
\centering
\begin{quantikz}[column sep=0.4cm]
    \lstick[wires=5]{\(n\)} & \ctrl{6}       & \ms{7} & \ctrl{5} & \qw            & \ctrl{5} & \rstick[wires=3]{\(\ceil{\frac{n}{2}}\)}\qw \\
                            & \ctrl{5}       & \qw    & \ctrl{4} & \qw            & \ctrl{4} & \qw \\
                            & \ctrl{4}       & \qw    & \ctrl{3} & \qw            & \ctrl{3} & \qw \\
                            & \ctrl{3}       & \qw    & \qw      & \ctrl{3}       & \qw      & \rstick[wires=3]{\(\floor{\frac{n}{2}} + 1\)}\qw \\
                            & \ctrl{2}       & \qw    & \qw      & \ctrl{2}       & \qw      & \qw \\
    \lstick{\(\ket{0}\)}    & \qw            & \qw    & \targ{}  & \ctrl{1}       & \targ{}  & \qw \\
                            & \gate{R_x} & \qw    & \qw      & \gate{R_x} & \qw      & \qw
\end{quantikz}
\caption{Splitting a multi-controlled gate with one zeroed ancilla into three multi-controlled gates with half the number of controls \cite{gidney:2015:1}.}
\label{fig:half}
\end{figure}
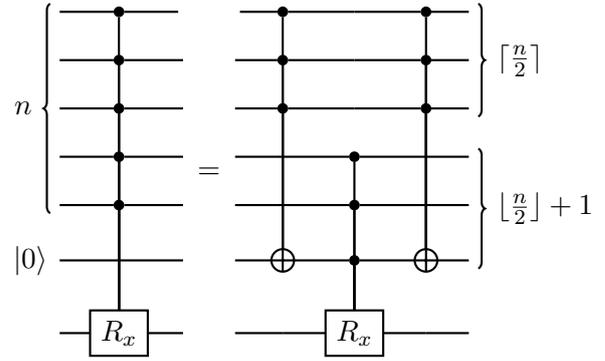

\begin{figure}[t]
    \centering
    \begin{quantikz}[column sep=0.33cm]
        & \ctrl{6} & \ms{7} & \qw      & \qw      & \ctrl{4} & \qw      & \qw      & \qw      & \ctrl{4} & \qw      & \qw \\
        & \ctrl{5} & \qw    & \qw      & \qw      & \ctrl{3} & \qw      & \qw      & \qw      & \ctrl{3} & \qw      & \qw \\
        & \ctrl{4} & \qw    & \qw      & \ctrl{3} & \qw      & \ctrl{3} & \qw      & \ctrl{3} & \qw      & \ctrl{3} & \qw \\
        & \ctrl{3} & \qw    & \ctrl{3} & \qw      & \qw      & \qw      & \ctrl{3} & \qw      & \qw      & \qw      & \qw \\
        & \qw      & \qw    & \qw      & \ctrl{1} & \targ{}  & \ctrl{1} & \qw      & \ctrl{1} & \targ{}  & \ctrl{1} & \qw \\
        & \qw      & \qw    & \ctrl{1} & \targ{}  & \qw      & \targ{}  & \ctrl{1} & \targ{}  & \qw      & \targ{}  & \qw \\
        & \targ{}  & \qw    & \targ{}  & \qw      & \qw      & \qw      & \targ{}  & \qw      & \qw      & \qw      & \qw
    \end{quantikz}
    \caption{Decomposition scheme for a multi-controlled \(X\) gate with access to two borrowed ancilla qubits that get returned to the state $\ket{0}$ \cite{barenco1995elementary}.}
    \label{fig:cnot-into-toffoli}
\end{figure}
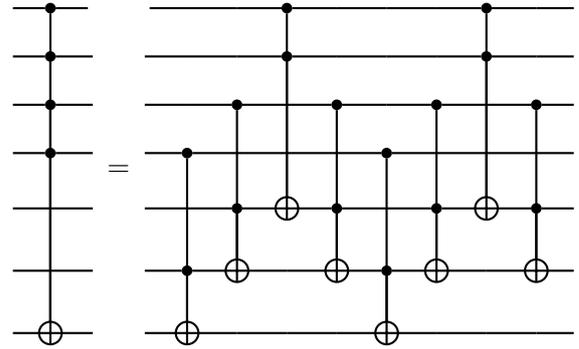

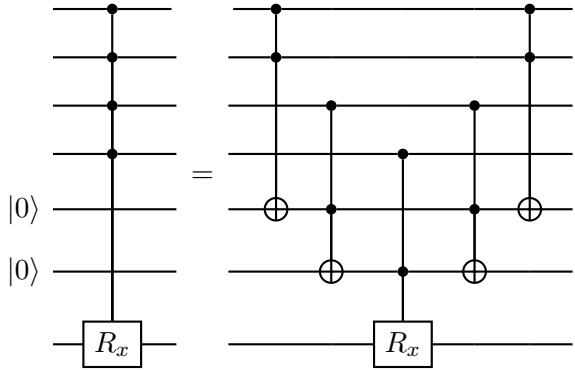
\begin{figure}[t]
    \centering
    \begin{quantikz}[column sep=0.4cm]
                             & \ctrl{6}    & \ms{7} & \ctrl{4} & \qw      & \qw        & \qw      & \ctrl{4} & \qw \\
                             & \ctrl{5}    & \qw    & \ctrl{3} & \qw      & \qw        & \qw      & \ctrl{3} & \qw \\
                             & \ctrl{4}    & \qw    & \qw      & \ctrl{3} & \qw        & \ctrl{3} & \qw      & \qw \\
                             & \ctrl{3}    & \qw    & \qw      & \qw      & \ctrl{3}   & \qw      & \qw      & \qw \\
        \lstick{\(\ket{0}\)} & \qw         & \qw    & \targ{}  & \ctrl{1} & \qw        & \ctrl{1} & \targ{}  & \qw \\
        \lstick{\(\ket{0}\)} & \qw         & \qw    & \qw      & \targ{}  & \ctrl{1}   & \targ{}  & \qw      & \qw \\
                             & \gate{R_x}  & \qw    & \qw      & \qw      & \gate{R_x} & \qw      & \qw      & \qw
    \end{quantikz}
    \caption{Decomposing a multi-controlled \(R_x \in SU(2)\) gate with access to two zeroed ancilla qubits.} %\textcolor{green}{@Nick - is there a citation for this decomposition? It might not need one because this seems to be the straightforward-naive method}.
    \label{fig:n-ancilla}
\end{figure}

\subsection{Gate counts for decompositions}
In all three of the variational algorithms we will focus on -- SA-QAOA, MA-QAOA, and DQVA -- the only multi-qubit entangling gate needed is the multi-controlled rotation gate, i.e. the \(C^n(R_x(\theta))\) gate, which we use to encode the constraints of the problem. For MIS, we use controlled rotations to encode information about the edges within the input graph.

In what follows, we use decomposition schemes from \cite{barenco1995elementary} and \cite{iten2016quantum} to give the total number of entangling gates required to decompose a \(C^n(R_x(\theta))\) gate. Since the rotation gate is a special unitary operator \(R_x(\theta) \in SU(2)\), we can make use of simpler decomposition schemes than would be possible for a general single-qubit gate \(U\). We will now give a general overview of the decomposition schemes used and show how the entangling gate counts were calculated.

\begin{table*}
    \centering
    \begin{threeparttable}
        \renewcommand*{\arraystretch}{1.25}
        \setlength{\tabcolsep}{0.75cm}
        \begin{tabular}{c c c c c c}
             \toprule
             & \multicolumn{2}{c}{\textbf{One ancilla}} & \multicolumn{2}{c}{\textbf{\(n\) ancillae}} & \textbf{No ancilla} \\
             & \(\mcS^2_2\) & \(\mcS^2_3\) & \(\mcS^2_2\) & \(\mcS^2_3\) & \(\mcS^3_2\) \\
             \midrule
             \(n = 1\) & 2 & 2 & 2 & 2 & 2 \\
             \(n = 2\) & 6 & 2 & 6 & 2 & 4 \\
             \(n = 3\) & 18 & 4 & 18 & 4 & 14 \\
             \(n = 4\) & 42 & 10 & 24 & 6 & 16 \\
             \(n = 5\) & 72 & 16 & 30 & 8 & 26 \\
             \midrule
             \(\vdots\) & \(16n-8\) & \(8n-24\) & \(6n\) & \(2n - 2\) & \(6n - 4\) ($n$ odd) \\
             & \multicolumn{2}{c}{\((n \geq 5)\)} & \multicolumn{2}{c}{\((n \geq 3)\)} & \(6n - 8\) ($n$ even) \\
             \bottomrule
        \end{tabular}
    \end{threeparttable}
    \caption{Exact entangling gate counts for a \(C^n(R_x)\) gate given access to zeroed ancilla using the decomposition schemes described in Section 3.}
    \label{tab:counts}
\end{table*}

\subsubsection{Decompositions with one ancilla}
Our first step in the decomposition process will be to split the single multi-control gate with \(n\) controls into three multi-control gates with \(n/2\) controls, following the zeroed bit construction from \cite{gidney:2015:1}. This step is illustrated for \(n = 5\) in Figure~\ref{fig:half}.

Next, as was shown in Figure \ref{fig:partialmixer}, since \(R_x \in SU(2)\) we can split up the resulting \(C^{\floor{n/2}+1}(R_x)\) gate into two multi-controlled Toffolis and four single qubit gates following Lemma 5.1 of \cite{barenco1995elementary}.

Then, we can decompose the resulting \(C^{\ceil{n/2}}(X)\) gates into a series of Toffoli gates by borrowing ancillae from the set of qubits not participating in the current gate, as described in Lemma 7.2 of \cite{barenco1995elementary} and shown in Figure~\ref{fig:cnot-into-toffoli}.

From this decomposition we can derive the asymptotic entangling gate counts. First, we consider the case where we have access to native Toffoli gates (i.e. \(\mcS^2_3\)). Working backwards, the decomposition scheme shown in Figure~\ref{fig:cnot-into-toffoli} with \(k\) controls requires \(4k - 8\) Toffolis to implement, since there are 4 Toffoli gates for each of the \(k-2\) controls. Then the decomposition in Figure~\ref{fig:half} followed by the decomposition in Figure~\ref{fig:partialmixer} gives two Toffolis with \(\ceil{n/2}\) controls and two Toffolis with \(\floor{n/2} + 1\) controls, so in total the number of entangling gates used is
\begin{eqnarray}
\begin{aligned}
    2\qty(4\ceil*{\frac{n}{2}} - 8) + 2&\qty(4\qty(\floor*{\frac{n}{2}} + 1) - 8) \\
    & = 8\qty(n + 1) - 32 \\
    & = 8n - 24,
\end{aligned}
\end{eqnarray}
for \(n \geq 5\). For \(n < 5\), there are too few controls for the decomposition scheme to work, but these cases can be calculated by hand. 

Next, we consider the case where we do not have access to native Toffoli gates (i.e. \(\mcS^2_2\)). The decomposition scheme shown in Figure~\ref{fig:cnot-into-toffoli} with \(k\) controls instead requires \(8k - 6\) \(\textsc{cnot}\)'s to implement (see Lemma 8 of \cite{iten2016quantum}) and then the same algebra as above gives an entangling gate count of \(16n - 8\), again for \(n \geq 5\). These results, along with the base cases of \(n < 5\) are shown in Table~\ref{tab:counts}.

\begin{table*}[t]
    \centering
    \begin{threeparttable}
        \renewcommand*{\arraystretch}{1.5}
        \begin{tabular}{c l c c}
            \toprule
            \textbf{Gate set}             & \textbf{\# of ancillae}      &
            \(C^{n}(R_x)\)                              &
            \(C^{n}(X)\)                                        \\
            \midrule
            \multirow{3}{*}{\(\mcS_2^2\)} & No ancillae & {\small\((28n - 24, 28n - 60)\)}              & ---        
            \\
            & One ancilla                  & {\small\((16n + 20, 16n
                - 6)\)}                 & {\small\((8n + 8, 8n - 4)\)}                          \\
            & \(n\) ancillae               & {\small\((8n - 8, 6n -
                6)\)}                             & ---                                                        \\
            \midrule
            \multirow{3}{*}{\(\mcS_3^2\)} & No ancillae & {\small\((12, 6, 14n - 38)\)}               & ---            
            \\
            & One ancilla                  & {\small\((4, 2, 8n -
                24)\)}                              & {\small\((0, 0, 4n - 12)\)}                           \\
            & \(n\) ancillae               & {\small\((6, 2, n-2)\)}
            & {\small\((0, 0, n-1)\)}                               \\
            \midrule
            \multirow{3}{*}{\(\mcS_m^2\)} & No ancillae                  &
            {\small\(\ca\qty(\frac{16}{m-2})n\) \(\mcnot{m-1}\)'s} & ---                    
            \\
            & One ancilla                  & {\small\(\ca\qty(\frac{8}{m-2})n\)
                \(\mcnot{m-1}\)'s}    & {\small\(\ca\qty(\frac{4}{m-2})n\) \(\mcnot{m-1}\)'s} \\
            & \(n\) ancillae               & {\small\(\ca\qty(\frac{1}{m-2})n\)
                \(\mcnot{m-1}\)'s}    & {\small\(\ca\qty(\frac{1}{m-2})n\) \(\mcnot{m-1}\)'s} \\
            \midrule
            \(\mcS_2^3\)                  & No ancillae                  & {\small\((10n - 10, 6n - 4)\)}         
            & ---                                                        \\
            \bottomrule
        \end{tabular}
    \end{threeparttable}
    \caption{
        Asymptotic gate counts for decompositions into different gate sets with differing amounts of available ancillae. The \(i\)-th entry in the ordered tuple shows how many \(i\)-qudit basis gates are required in \(\mcS^d_m\) to decompose that type of gate (e.g. \((a,b)\) would require \(a\) single qubit gates and \(b\) \textsc{cnot}'s). Cells with no entry have no better decomposition than the equivalent \(C^n(R_x)\) decomposition in the same row. For schemes with gate counts that depend on the parity of \(n\), we simply report the larger gate count. Note that the asymptotic counts reported here consider \textit{burnable} ancillae and may differ from the counts given in Table \ref{tab:counts} which considered \textit{zeroed} ancillae. The derivation of the gate counts is contained in Appendix \ref{app:asymptotic_counts}.
        }
    \label{tab:the-spreadsheet}
\end{table*}

\begin{table*}[t]
    \centering
    \begin{threeparttable}
        \renewcommand*{\arraystretch}{1.25}
        \setlength{\tabcolsep}{0.28cm}
        \small
        \begin{tabular}{l c c c c c c c}
            \toprule
            \textbf{Threshold} & \textbf{Requirement} &
            \mc{\thead{\small\(\mcF_1 = 95\%\)\\\small\(\mcF_2 = 90\%\)}} &
            \mc{\thead{\small\(\mcF_1 = 99\%\)\\\small\(\mcF_2 = 95\%\)}} &
            \mc{\thead{\small\(\mcF_1 = 99.9\%\)\\\small\(\mcF_2 = 99\%\)}} &
            \mc{\thead{\small\(\mcF_1 = 99.99\%\)\\\small\(\mcF_2 = 99.9\%\)}} &
            \mc{\thead{\small\(\mcF_1 = 99.999\%\)\\\small\(\mcF_2 = 99.99\%\)}} \\
            \midrule
            \(\mcS_3^2 \succ \mcS_2^2\) & \(\mcF_3 > \mcF_1^2 \mcF_2^2\) & \(73\%\) & \(88\%\) & \(97.8\%\) & \(99.78\%\) & \(99.978\%\) \\
            \(\mcS_4^2 \succ \mcS_3^2\) & \(\mcF_4 > \mcF_3^2\) & \(53\%\) & \(78\%\) & \(95.7\%\) & \(99.56\%\) & \(99.956\%\) \\
            \(\mcS_5^2 \succ \mcS_4^2\) & \(\mcF_5^2 > \mcF_4^3\)  & \(39\%\) & \(69\%\) & \(93.6\%\) & \(99.34\%\) & \(99.934\%\) \\
            \(\mcS_6^2 \succ \mcS_5^2\) & \(\mcF_6^3 > \mcF_5^4\) & \(29\%\) & \(61\%\) & \(91.2\%\) & \(99.12\%\) & \(99.912\%\) \\
            \(\mcS_7^2 \succ \mcS_6^2\) & \(\mcF_7^4 > \mcF_6^5\) & \(21\%\) & \(54\%\) & \(89.6\%\) & \(98.91\%\) & \(99.890\%\) \\
            \(\mcS_8^2 \succ \mcS_7^2\) & \(\mcF_8^5 > \mcF_7^6\)  & \(15\%\) & \(48\%\) & \(87.6\%\) & \(98.69\%\) & \(99.868\%\) \\
            \bottomrule
        \end{tabular}
    \end{threeparttable}
    \caption{
        Summary of gate set thresholds, their gate fidelity requirements,
        and the required largest gate fidelity under various assumptions for \(\mcF_1\) and \(\mcF_2\), the one- and two-qubit gate fidelities, respectively. For larger gate sets, we assume that each of the smaller gates just meet the fidelity threshold (for example, to calculate the threshold for \(\mcF_5\), we assume recursively that \(\mcF_3\) and \(\mcF_4\) equal their respective threshold values). The center column with values of \(\mcF_1 \approx 99.9\%\) and \(\mcF_2 \approx 99\%\) is about where current NISQ technologies stand \cite{tomesh2022supermarq, google2023suppressing}.}
    \label{tab:threshold-results}
\end{table*}

\subsubsection{Decompositions with $n$ ancillae}

If we have access to many ancilla qubits -- i.e. roughly the same number as there are controls -- then we can use an even simpler decomposition scheme, where we decompose directly into 2-controlled Toffoli gates. The scheme is similar to that shown in Figure~\ref{fig:cnot-into-toffoli}, and is shown in Figure~\ref{fig:n-ancilla}.

We can then decompose the \(C^2(R_x)\) gate in Figure~\ref{fig:n-ancilla} into four single qubit gates and two Toffoli gates.
%We can then apply the decomposition in Figure~\ref{fig:n-ancilla} to the resulting \(C^2(R_x)\) gate. 
Then, directly counting the number of Toffoli gates used for the \(\mcS^2_3\) count, there are two Toffoli gates for each of the \(n - 1\) controls, giving \(2n - 2\) Toffoli gates required for \(n \geq 3\). To get the \(\mcS^2_2\) count, we use the tricks shown in the proof of Lemma 8 of \cite{iten2016quantum} to lower the number of \(\textsc{cnot}\)'s required to only \(6n\) for \(n \geq 3\). These results, along with the base cases, are also shown in Table~\ref{tab:counts}.

\subsubsection{Additional counts}
Using similar methodology to that shown above, one can derive exact gate counts for decomposing such gates with access to burnable ancilla -- meaning that the ancilla qubits are initialized to \(\ket{0}\) and do not need to be returned to this state -- we have carried this analysis out, following the decomposition schemes given in \cite{barenco1995elementary, gidney:2015:1, gidney:2015:2, gidney:2015:3}, with the results summarized in Table~\ref{tab:the-spreadsheet}. The derivations of the gate counts can be found in Appendix \ref{app:asymptotic_counts} as they may prove useful for future reference.

We also include the asymptotic gate counts for the $\mcS^3_2$ gate set, which includes single- and two-qutrit gates, based on the decomposition schemes given in \cite{gokhale2019asymptotic}. These decompositions require no ancillae but we include them in our comparisons in Section \ref{resource} as an example of an interesting tradeoff between resource efficiency and the complexity of physical implementation. 
Intuitively, the qutrits (with basis states $\ket{0}, \ket{1}, \text{and} \ket{2}$) can use the extra $\ket{2}$ state as an efficient scratch space to store parity information.
This enables the \(\mcS^3_2\) gate set to achieve the same asymptotic cost as the \(\mcS^2_2\) gate set \textit{with n ancillae} when decomposing multi-controlled $R_x$ and $X$ gates.

\subsection{Fidelity thresholds}

We can also use the results from Table~\ref{tab:the-spreadsheet} to derive gate set thresholds -- i.e. under what conditions it is asymptotically beneficial to decompose into one gate set over another. For example, if the \(i\)-qubit basis gate of \(\mcS_m\) has gate fidelity \(0 \leq \mcF_i \leq 1\) (e.g. \(\mcF_2\) is the \textsc{cnot} fidelity, etc.), then by comparing the one ancilla decompositions for \(\mcS_2^2\) and \(\mcS_3^2\) in Table~\ref{tab:the-spreadsheet}, at large \(n\) the process fidelity for a \(C^n(R_x)\) gate in \(\mcS_2^2\) is \(\mcF_1^{16n}\mcF_2^{16n}\) while the process fidelity in \(\mcS_3^2\) is \(\mcF_3^{8n}\). In order for \(\mcS_3^2\) to be better than \(\mcS_2^2\), we must have
\[
    \mcF_3^{8n} > \mcF_1^{16n}\mcF_2^{16n} \implies \mcF_3 > \mcF_1^2 \mcF_2^2.
\]
In other words, in order for it to be asymptotically beneficial to stop decompositions early using native Toffolis, the Toffoli fidelity must be greater than the product of the squares of the single qubit and \(\textsc{cnot}\) fidelities.

A summary of these results, as well as some experimental targets for various gate fidelity estimates, is given in Table~\ref{tab:threshold-results}. The central column, where \(\mcF_1 \approx 99.9\%\) and \(\mcF_2 \approx 99\%\), most closely represents the capabilities of many current NISQ computers \cite{tomesh2022supermarq, google2023suppressing}. Using these systems as an example, if they natively supported a three-qubit gate with fidelity \(\mcF_3 > 97.8\%\), it would be asymptotically advantageous to decompose large multi-controlled gates into \(\mcS_3^2\) instead of \(\mcS_2^2\). This agrees with other theoretical results and simulations that suggest that such gate set optimizations would be possible, for example in \cite{gulliksen:2015}.

\section{Analyzing Classical and Quantum Resource Tradeoffs}\label{resource}

Hybrid quantum-classical algorithms offer architectural tradeoffs which allow both the CPU and QPU to share the burden of a computation.
For a given instance of a target problem (e.g., solving MIS on a specific graph) the exact value of these tradeoffs that will maximize performance depends on the characteristics of the specific algorithm and hardware that are used.
In this section we consider the impact that the number of variational parameters, number of rounds of optimization, and native gate set supported by the hardware backend have on the final performance, runtime, and circuit depth for three flavors of the QAOA.
We begin by examining the tradeoff between the size of the independent sets produced by the algorithms on a suite of benchmark graphs and the classical computational resources they require which includes the number of variational parameters and the number of rounds of classical optimization.
We then compare the number of quantum operations required by the different algorithms as a function of the native gate set supported by hardware. 

\subsection{Performance vs. Classical Resources}

\begin{figure*}[t]
\centering
    \includegraphics[width=\linewidth]{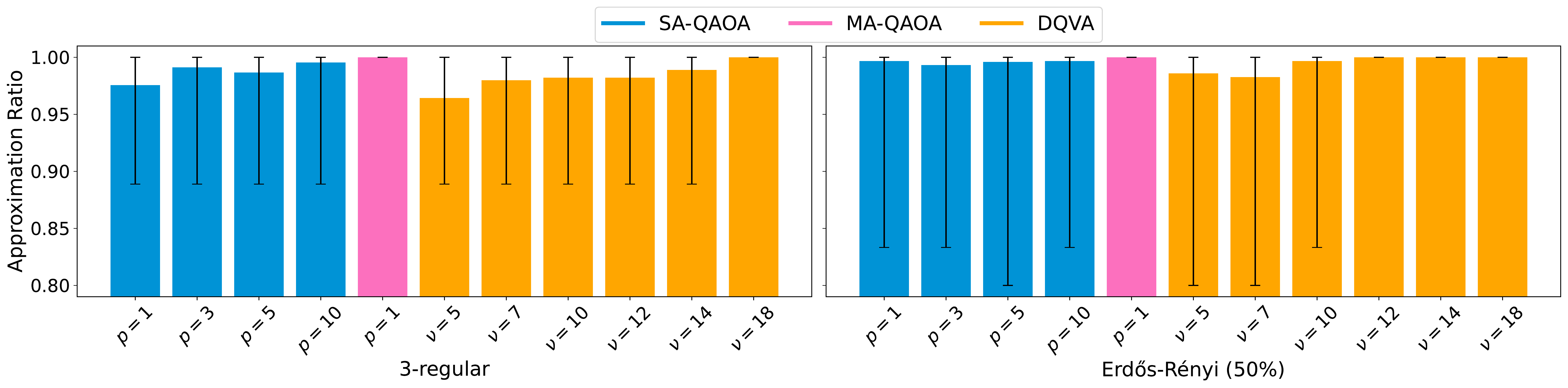}
    \caption{The average approximation ratio attained by each QAOA variant with respect to the optimal MIS over 50 20-node benchmark graphs. For each benchmark graph, every algorithm is executed 30 times and the best performing run is saved. The error bars show the maximum and minimum approximation ratio over the 50 benchmark graphs. Each algorithm is able to find the optimal solution at least once. This is even true for some algorithms when the number of parameters is much less than the size of the graph.}
    \label{fig:performance}
\end{figure*}

We consider three variations of the QAOA: SA-QAOA, MA-QAOA, and DQVA as defined in Section \ref{QAOA}. Figure \ref{fig:performance} shows the maximum, minimum, and average approximation ratios achieved by the three algorithms as a function of the number of classical parameters used within each variational ansatz. 
We simulated the execution of each algorithm on both 3-regular and Erd\"os-R\'enyi graphs containing 20 nodes. For the Erd\"os-R\'enyi graphs the probability of any edge existing between two nodes was set to 50\%. An open-source implementation of the code used to construct the quantum circuits and simulate the execution of the algorithms is available on Github \cite{tomesh2022github}.

The simulation results were collected over ensembles of 50 benchmark graphs and for each graph the best performing execution over 30 repetitions was saved. 
Prior work has shown that the number of trials needed to converge to good solutions can scale strongly with the number of paramters. For example, while studying parameter transfer for solving MaxCut problems with QAOA, \cite{shaydulin2022parameter} allocated 50, 200, and 1,500 trials for $p=1,2,3$ SA-QAOA, respectively.
In this work our aim is to analyze the tradeoff between performance -- as measured by the final approximation ratio -- and the classical resources required to execute the QAOA.
An equal-repetition comparison helps to capture this tradeoff and serves to represent the difficulty of optimizing ansatzes with an increasing number of classical parameters. Indeed, Figure \ref{fig:performance} shows that each algorithm is able attain comparably high approximation ratios so long as the number of classical parameters is not too small.

\begin{figure}[t]
    \includegraphics[width=\linewidth]{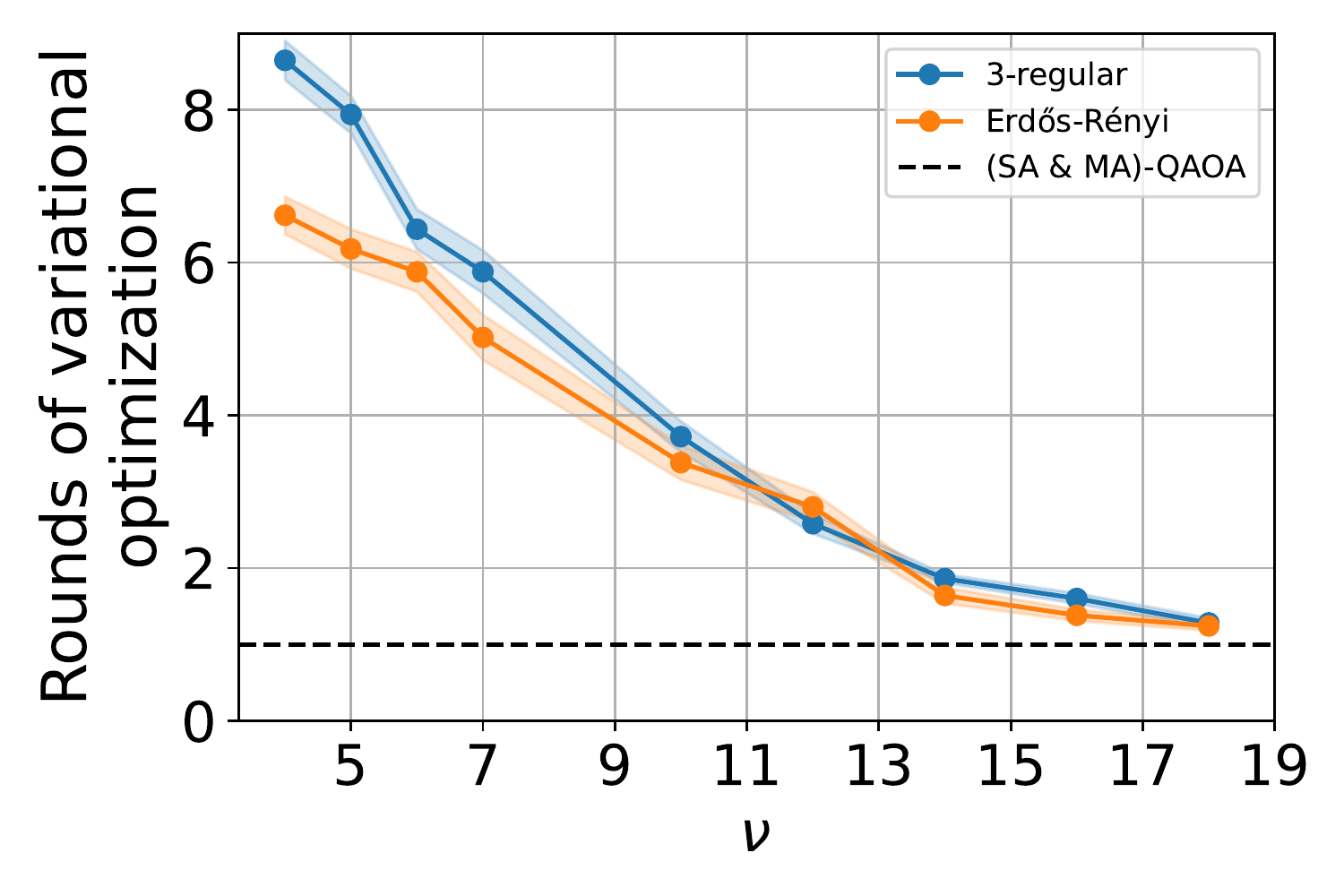}
    \caption{Average number of rounds of variational optimization incurred by the DQVA as a function of the number of parameters $\nu$ across both the 3-regular and Erd\"os-R\'enyi benchmark graphs. The shaded areas denote one standard error from the mean. DQVA makes a tradeoff between the classical and quantum resources needed to solve the MIS problem. Fewer parameters and partial mixers can be used at the cost of additional rounds of variational optimization. At $\nu \gtrsim 10$ the DQVA incurs $2-4$ rounds of optimization and finds approximation ratios comparable to the SA-QAOA and MA-QAOA.}
    \label{fig:convergence}
\end{figure}

Even though each algorithm is evaluated on a target graph for an equal number of trials, the classical resources required within a single trial is also variable. This is qualitatively shown in Table \ref{tab:overview} where both the SA-QAOA and MA-QAOA require a single round of classical optimization but the DQVA can require multiple.
Figure \ref{fig:convergence} quantitatively illustrates this point by showing the average number of rounds of classical optimization required until the DQVA reaches its final solution. 
Instances of the DQVA which use few parameters, $\nu < \frac{n}{2}$, require many additional rounds of optimization (between 4 to 9) and on average they find independent sets smaller than what the SA-QAOA and MA-QAOA are able to find in a single round of optimization as seen in Figure \ref{fig:performance}.
However, for larger values of $\nu \gtrsim \frac{n}{2}$ comparable performance can be achieved with approximately 1 to 3 additional rounds of optimization. The DQVA balances this additional cost in classical resources by requiring quantum circuits which contain many fewer gates than the SA-QAOA and MA-QAOA as discussed below.

\subsection{Native Gate Set vs. Quantum Resources}

%\newgeometry{margin=1cm} % modify this if you need even more space
%\begin{landscape}

\begin{figure*}[t]
    \centering
    \includegraphics[width=0.8\linewidth]{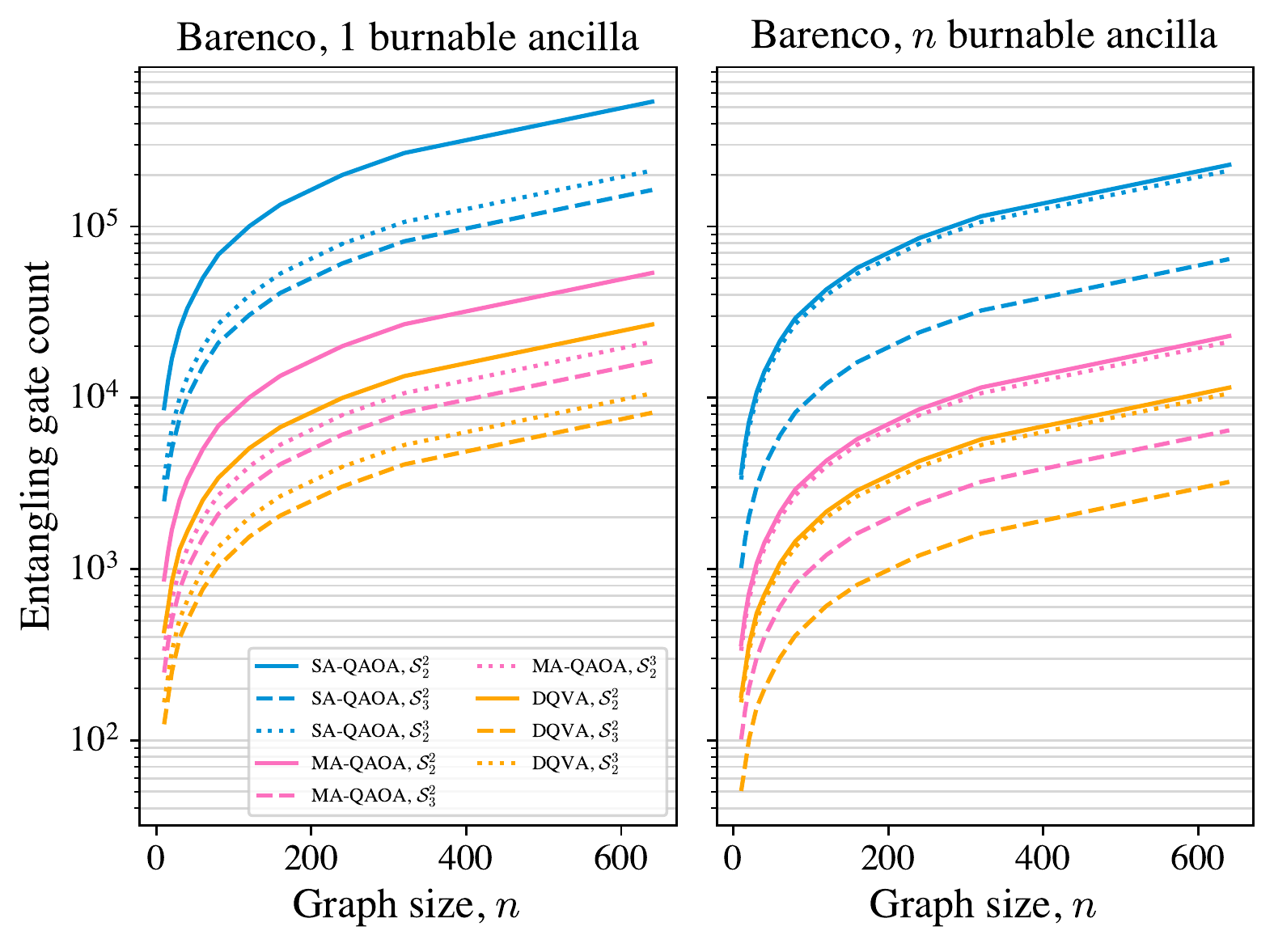}
    \caption{Total entangling gate count on Erd\"os-R\'enyi graphs with constant average density $d=6$ (i.e., the edge probability is always set to $\frac{d}{m-1}$) for the $p=10$ SA-QAOA (blue) with 20 variational parameters, $p=1$ MA-QAOA (pink) with $m$ variational parameters, and $\nu=\frac{m}{2}$ DQVA (orange). We compare the gate costs incurred by decomposing the ansatz into the $\mathcal{S}^2_2$ (solid) and $\mathcal{S}^2_3$ (dashed) gate sets with access to $1$ and $n$ burnable ancillae. We also include the gate overhead associated with decompositions targeting the $S^3_2$ (dotted) qutrit gate set. Note that the decompositions in this gate set require no ancilla and the same counts are shown in both panels.
    The $S^2_3$ gate set always produces the most efficient decompositions across all graph sizes. The plot on the right highlights the resource efficiency of qutrits --- the $S^3_2$ gate set incurs a slightly smaller gate cost than the $S^2_2$ qubit gate set while requiring only half as many qutrits since its decomposition requires no ancilla. 
    }
    \label{fig:numerical_gate_counts}
\end{figure*}

The exact gate counts for the benchmark graphs used in Figure \ref{fig:performance} are contained in Appendix \ref{app:numerical_counts}.
To better understand the circuit sizes that would be encountered in more realistic workloads targeting larger graphs, we used the exact gate counts given in Table \ref{tab:counts} to compute the number of entangling gates required across algorithms and native gate sets.
Since variational circuit simulations quickly become intractable beyond $20$ to $30$ qubits, we were unable to verify the values of $p$ and $\nu$ needed by SA-QAOA, MA-QAOA, and the DQVA, respectively, to ensure adequate performance as the graph size increased.
Instead, we selected the $p=10$ SA-QAOA, $p=1$ MA-QAOA, and $\nu=\frac{n}{2}$ DQVA which showed comparable performance on the 20-node Erd\"os-R\'enyi graphs in Figure \ref{fig:performance} and calculated the gate counts for these ansatzes when targeting average degree $d=6$ Erd\"os-R\'enyi graphs with up to 640 nodes.
Figure \ref{fig:numerical_gate_counts} shows the final number of entangling gates of each ansatz for the $S^2_2$, $S^2_3$, and $S^3_2$ gate sets.
Additionally we report the cost for separate scenarios where only a single ancilla or $n$ ancilla are available. Note that this only impacts the qubit ($S^2_2$ and $S^2_3$) decompositions because the qutrit $S^3_2$ decomposition requires no extra ancilla. In the future, we expect the $n$ ancilla case to be more common as quantum processors scale to larger sizes since it is unlikely that every program will require 100\% of the qubits available on a device \cite{gokhale2019asymptotic}.

The impact of the constant factors in front of the asymptotic scaling given in Tables \ref{tab:counts} and \ref{tab:the-spreadsheet} appear clearly in Figure \ref{fig:numerical_gate_counts}. The $S^2_3$ gate set consistently produces the shortest circuits across graph sizes and amount of available ancillae. Interestingly, the gate counts for the $S^2_2$, with $n$ ancillae, and $S^3_2$, with no ancilla, decompositions are nearly equivalent. 

\begin{figure*}[t]
    \centering
    \includegraphics[width=0.9\linewidth]{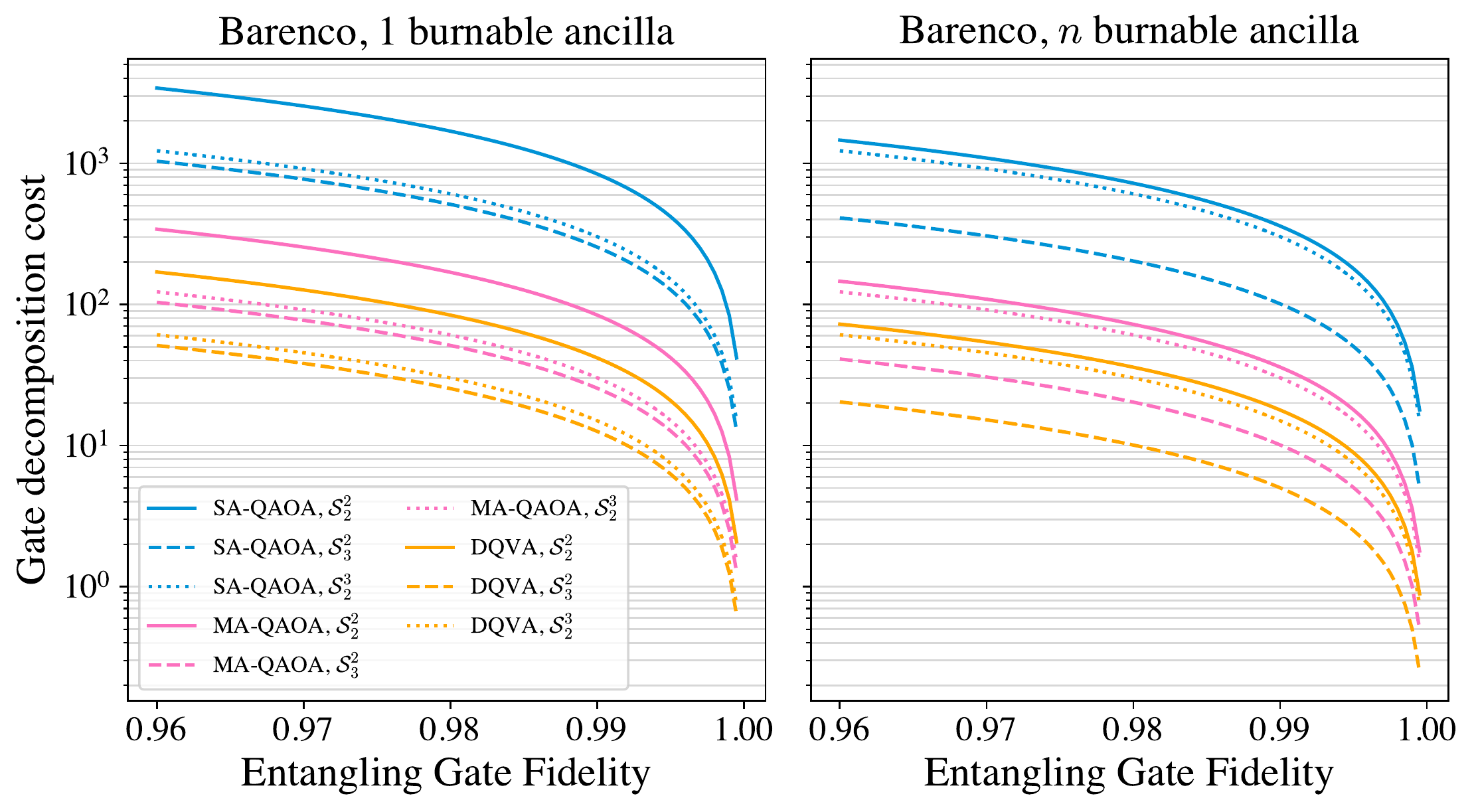}
    \caption{Gate decomposition cost (GDC) for the QAOA variants as a function of entangling gate fidelity. These values were computed using 100-node Erd\"os-R\'enyi graphs with an average density $d=6$. The more efficient decompositions produced by the $\mcS^2_3$ and $\mcS^3_2$ gate sets allow for lower fidelity entangling gates while maintaining an equivalent GDC as compared to the $\mcS^2_2$ decompositions.}
    \label{fig:decomp-infid-score}
\end{figure*}

%\begin{table}[h]
%\begin{tabular}{cccc}
%\hline
%\textbf{Circuit} & \textbf{Qiskit} & \textbf{tket} & %\textbf{this work} \\ \hline
%QAOA p=10        & a                              & b             & c                  \\ \hline
%MA-QAOA p=1      & d                              & e             & f                  \\ \hline
%DQVA $\nu = 10$  & g                              & h             & I                  \\ \hline
%\end{tabular}
%\end{table}
%\textcolor{red}{One final experiment to present: create a table for some benchmark circuits and the cost of their decomposition using (1) the techniques in our paper, (2) using the IBM compiler with optimization level 3, (3) using tket. TURN THIS INTO A FIGURE}
    
\subsection{Decomposition Cost}
In order to compare the usefulness of the different decomposition methods, we define a metric in Equation \eqref{eqn:dis} which we call the Gate Decomposition Cost (GDC). 
\begin{equation}\label{eqn:dis}
    \text{GDC} \coloneqq -\log\left({\prod_i^{\text{Types}}} \mcF_i^{n_i}\right)
\end{equation}
The GDC takes into account the number, $n_i$, and type (e.g., $C^1X$, $C^2X$, etc.) of basis gates that we decompose the large multi-controlled gates into and also takes into account their respective fidelities, $\mcF_i$.

This is a good metric that is generalizable beyond this paper since gate fidelity products are an effective approach in comparing the output of the quantum circuit to the ideal output when fidelities are all 1. Even though gate fidelities are not invariant for different physical setups, the GDC will always decrease for smaller fidelities and increase for larger ones. The costs and benefits of each decomposition will always be captured by the product in the logarithmic function. We can similarly define a Gate Decomposition Score (GDS) as 1-GDC to compare how well one circuit helps solve the MIS problem to another one with a different decomposition.

Figure \ref{fig:decomp-infid-score} shows the average GDC of the different QAOA variants when targeting 100-node Erd\"os-R\'enyi graphs. Note that the gate fidelity given along the X-axis is assumed for every entangling operation within the gate set, however, in practice, we expect two-qubit gates to have higher fidelities than three-qubit gates for most hardware platforms. This assumption mainly impacts the GDC for the $\mcS^2_3$ gate set which produces both two- and three-qubit gates in the final circuit decompositions, but we expect the impact to be minimal as we observed that the number of three-qubit gates always dominated the two-qubit gate count by a factor of approximately 1000.

The GDC provides a useful means of combining the gate decomposition results with the gate fidelities we expect to see in future hardware. Comparing the different gate set decompositions of a particular QAOA variant along equal GDC lines indicates the easing fidelity requirements afforded by lower basis gate overheads. These numerical results confirm the asymptotic values provided in Table \ref{tab:threshold-results} and reflect the diminishing gains won by the $\mcS^2_3$ gate set as the quality of the two-qubit gates approaches unity.

%%%%%%%%%%%%%%%%%%%%%%%%%%%%%%%%%%%%%%%%%%%%%%%%%%%%%%%%%%%%
%\section{Conclusion and Future Directions}\label{conclusions}
\section{Conclusions}\label{conclusions}

%\textcolor{red}{This paper serves as a guidepost for future system design. We give system designers a target fidelity to shoot for. One reviewer said he appreciated we didn't make grandiose claims about QAOA but asked us to give more context on how that algorithm + this work would have more impact.}

% Summary of results
Hybrid variational algorithms are a promising means for extending the capabilities of classical computation with noisy, intermediate-scale, quantum resources. 
While the practical utility of heuristic approaches such as the QAOA remains an open question, recent work, both theoretical \cite{basso2021quantum} and empirical \cite{shaydulin2023evidence}, has provided interesting avenues for future investigation.
For these applications, we consider tradeoffs at an algorithmic level, which exchange fewer quantum resources for additional classical computation. Concretely, we consider variants of the QAOA applied to the MIS problem and study how increasing or decreasing the number of parameters --- which is proportional to the depth of the quantum circuits --- affects performance. We find that the MA-QAOA, which parameterizes each individual partial mixer, outperforms the SA-QAOA while requiring much shorter quantum circuits. 
Furthermore, the DQVA extends this idea by allowing some of those individual parameters to be turned off. This enables the DQVA to further reduce the quantum cost compared to the MA-QAOA at the expense of additional iterations of classical optimization.

In addition to these algorithmic tradeoffs, we also consider the effect that the hardware's native gate set has on the overhead incurred during gate decomposition. We analyze the gate decomposition schemes introduced in prior work \cite{barenco1995elementary, gokhale2019asymptotic}, and provide asymptotic as well as numerical gate counts specific to the multi-controlled $R_x$ rotations which are ubiquitous in constrained QAOA for MIS. While our asymptotic gate counts show that all of the decompositions have the same $\mathcal{O}(n)$ scaling, our numerical calculations reveal the practical impact of the constant scaling factors as the size of the problem graph grows. For example, for the 600-node graphs, the gate counts for the same QAOA circuit under different decompositions may differ by as much as $\num{1.6e5}$ entangling gates.

The asymptotic gate counts reported in Section \ref{decompositions} show the advantages, in terms of the gate decomposition overhead, of augmenting a quantum computer's typical single- and two-qubit gate set with higher-degree native gates as well as additional qudit states.
This is reinforced by the numerical simulations in Section \ref{resource} which suggest that the gains in decomposition efficiency can outweigh the effects of lower-fidelity three-qubit and two-qutrit gates.

% Application of these results to local search and divide and conquer algorithms, adapt-qaoa, adapt-vqe

However, our analysis of the tradeoffs in the decomposition overheads is not specific to this task and is applicable to many other quantum algorithms which rely on these gate operations including adders \cite{draper2004logarithmic, chakrabarti2008designing}, variational quantum eigensolvers with particle-conserving unitaries \cite{arrazola2022universal}, and measurement-free error correction \cite{crow2016improved, premakumar2020measurement}.

The trainability across the three QAOA variants exhibits relatively minor variations. The primary distinction, and a focal point of our research, lies in the total count of quantum gates and optimization rounds required. Our work primarily concentrates on enhancing the implementation of these algorithms by furnishing system designers with clear fidelity targets. This aids in discerning the advantages of supporting more intricate gate operations, particularly those involving many-body or higher-dimensional scenarios. These contributions are seen as complementary to ongoing initiatives addressing trainability issues in Variational Quantum Algorithms (VQAs). By elevating hardware fidelity and lowering the GDCs, we anticipate a reduction in the overall number of shots required to measure the relevant operator for other optimization schemes, be it the gradient or any other operator.

Constrained combinatorial optimization problems, and particularly MIS, have widespread applications including portfolio optimization in finance \cite{kalra2018portfolio}, the logistics of job scheduling \cite{lawler1973optimal}, and the modeling of genomics data \cite{auyeung2005new}. Obtaining accurate solutions to these problems in a time and cost efficient manner is a significant challenge. Many quantum-classical algorithms have been developed to target these problems, and they are often inspired by classical approaches such as divide and conquer and local search \cite{saleem2021quantum, tomesh2022quantum}. The analysis presented here is complementary with these algorithmic approaches as they use the same basic quantum gates.

\section{Future Directions}\label{future}
% Potential for hardware executions
Hardware experiments demonstrating the performance of constrained-QAOA are especially exciting. The crux of constrained-QAOA is the subspace-restricted evolution within the set of feasible solutions. Therefore, it is critically important to study the noise effects when running on real quantum hardware and developing mitigation methods. New hardware is scheduled to be made available which natively supports multi-qubit gates and we plan to experimentally test these algorithms.

More recently, some architectures have proposed methods to natively support highly-controlled gates in the quantum instruction set architecture. However, as the number of controls increases, these native gates will experience lower fidelity due to the increased experimental complexity. In neutral atom quantum computers, these gates can be implemented using Rydberg interactions~\cite{isenhower:2011,saffman2016quantum}. In fact, using these native Rydberg gates allows for more efficient implementations of Grover's search algorithm~\cite{molmer:2011,petrosyan:2016}, and others have shown via quantum Monte Carlo simulations that native Toffoli gates implemented via Rydberg blockades may have higher fidelity than the decomposition into one- and two-qubit Rydberg gates \cite{gulliksen:2015}, in spite of the Toffoli's naturally lower fidelity. Using numerical simulations and error analysis, error rates of \(1 - \mcF < 10\%\) for \(k = 35\)~\cite{isenhower:2011} and \(1 - \mcF < 1\%\) for \(k = 20\)~\cite{khazali:2020} are possible using these Rydberg gates.

Highly-controlled gates can also be implemented using Ising interactions similar to the two-qubit M{\o}lmer-S{\o}rensen gates typically used in trapped ion systems \cite{espinoza:2021,katz:2022}.
These gates are based on a single-step implementation protocol that scales with the number of qubits \cite{rasmussen:2020}. They demonstrate high fidelity (\(\mcF > 99\%\)) gates on 3 to 7 qubits  with gate times \(< \SI{1}{ms}\)~\cite{espinoza:2021}. 

Various approaches to natively implement highly-controlled gates also exist in superconducting systems. One approach involves using adiabatic evolution of excitation-exchange eigenstates, similar to those used in Rydberg atoms~\cite{khazali:2020}. Other demonstrations have relied on a driven Ising-type model similar to those used in trapped ions~\cite{rasmussen:2020}, and the native cross-resonance effect on fixed-frequency transmons \cite{kim2022high}. 
The new developments and experimental demonstrations of natively supported multi-controlled gates presents an exciting opportunity for implementing and testing algorithms such as the constrained QAOA on near-term quantum computers.

%%%%%%%%%%%%%%%%%%%%%%%%%%%%%%%%%%%%%%%%%%%%%%%%%%%%%%%%%%%%
\section{Acknowledgments}
T.T. and N.A. is supported in part by EPiQC, an NSF Expedition in Computing, under grant CCF-1730082. Z.S. is supported by the National Science Foundation under Award No. 2037984, and QNEXT.
 
 The submitted manuscript has been created by UChicago Argonne, LLC, Operator of Argonne National Laboratory (``Argonne”). Argonne, a U.S. Department of Energy Office of Science laboratory, is operated under Contract No. DE-AC02-06CH11357. The U.S. Government retains for itself, and others acting on its behalf, a paid-up nonexclusive, irrevocable worldwide license in said article to reproduce, prepare derivative works, distribute copies to the public, and perform publicly and display publicly, by or on behalf of the Government. The Department of Energy will provide public access to these results of federally sponsored research in accordance with the DOE Public Access Plan (\url{http://energy.gov/downloads/doe-public-access-plan}).

%%%%%%%%%%%%%%%%%%%%%%%%%%%%%%%%%%%%%%%%%%%%%%%%%%%%%%%%%%%%
\def\bibsection{\section{References}}
\bibliographystyle{unsrtnat}
\bibliography{main}

%%%%%%%%%%%%%%%%%%%%%%%%%%%%%%%%%%%%%%%%%%%%%%%%%%%%%%%%%%%%

\onecolumn
\newpage
\appendix

%\section{Derivation of Asymptotic Gate Counts}\label{app:asymptotic_counts}
%This section describes how the entries in \ref{tab:the-spreadsheet} were calculated.

\section{Gate Counts for Benchmark Graphs}\label{app:numerical_counts}
\begin{table*}[h!]
\centering
\tabcolsep=0.1cm
\resizebox{\linewidth}{!}{%
\begin{threeparttable}
    \begin{tabular}{ccccccccccccccc}
    \toprule
    \multicolumn{1}{c}{}        & \multicolumn{13}{c}{Average $C^k(R_x(\theta))$ gate counts for 20 node graphs}                                                                          &            \\
    \multicolumn{1}{c}{Circuit} & $C^3(R_x)$    & $C^4(R_x)$ & $C^5(R_x)$ & $C^6(R_x)$ & $C^7(R_x)$ & $C^8(R_x)$ & $C^9(R_x)$ & $C^{10}(R_x)$ & $C^{11}(R_x)$ & $C^{12}(R_x)$ & $C^{13}(R_x)$ & $C^{14}(R_x)$ & $C^{15}(R_x)$ & $C^{16}(R_x)$ \\
    \midrule
    SA-QAOA $p=1$                  & (20, 0.08) & 0.16    & 0.38    & 0.94    & 2.14    & 2.9     & 3.74    & 3.7        & 2.84       & 1.62       & 0.96       & 0.42       & 0.06       & 0.06       \\
    SA-QAOA $p=5$                  & (100, 0.4) & 0.8     & 1.9     & 4.7     & 10.7    & 14.5    & 18.7    & 18.5       & 14.2       & 8.1        & 4.8        & 2.1        & 0.3        & 0.3        \\
    SA-QAOA $p=10$                 & (200, 0.8) & 1.6     & 3.8     & 9.4     & 21.4    & 29.0    & 37.4    & 37.0       & 28.4       & 16.2       & 9.6        & 4.2        & 0.6        & 0.6        \\
    \midrule
    MA-QAOA $p=1$                  & (20, 0.08) & 0.16    & 0.38    & 0.94    & 2.14    & 2.9     & 3.74    & 3.7        & 2.84       & 1.62       & 0.96       & 0.42       & 0.06       & 0.06       \\
    \midrule
    DQVA $\nu=5$                     & (4, 0.01)  & 0.01    & 0.06    & 0.14    & 0.33    & 0.54    & 0.71    & 0.77       & 0.67       & 0.41       & 0.25       & 0.09       & 0.01       & 0.01       \\
    DQVA $\nu=7$                     & (6, 0.01)  & 0.03    & 0.07    & 0.17    & 0.45    & 0.87    & 1.15    & 1.2        & 0.92       & 0.56       & 0.36       & 0.17       & 0.02       & 0.03       \\
    DQVA $\nu=10$                    & (9, 0.01)  & 0.01    & 0.12    & 0.24    & 0.62    & 1.12    & 1.73    & 1.8        & 1.59       & 0.94       & 0.52       & 0.24       & 0.04       & 0.02       \\
    DQVA $\nu=12$                    & (11, 0.01) & 0.04    & 0.13    & 0.33    & 0.81    & 1.41    & 1.97    & 2.22       & 1.93       & 1.08       & 0.7        & 0.32       & 0.04       & 0.04       \\
    DQVA $\nu=14$                    & (13, 0.01) & 0.03    & 0.08    & 0.38    & 1.01    & 1.66    & 2.44    & 2.63       & 2.22       & 1.31       & 0.76       & 0.37       & 0.05       & 0.05       \\
    DQVA $\nu=18$                    & (17, 0.01) & 0.06    & 0.08    & 0.47    & 1.07    & 1.87    & 2.72    & 3.12       & 2.53       & 1.51       & 0.91       & 0.41       & 0.06       & 0.06       \\
    \bottomrule
    \end{tabular}
\end{threeparttable}
}
\caption{Average number of $C^k(R_x(\theta))$ gates that appeared in the variational circuits for SA-QAOA, MA-QAOA, and DQVA across all 20-node benchmark 3-regular and Erd\"os-R\'enyi graphs used to generate Figure \ref{fig:performance}. The quantum circuits targeting the 3-regular graphs are composed entirely of $C^3(R_x(\theta))$ gates so two values are listed in that column (3-regular counts on the left, Erd\"os-R\'enyi on the right). All other gate counts correspond to quantum circuits targeting the benchmark Erd\"os-R\'enyi graphs. Across both graph types and number of controls, the DQVA produces variational circuits requiring fewer quantum resources.}
\label{tab:20nodecounts}
\end{table*}

Table \ref{tab:20nodecounts} reports the number and kind of partial mixers that appear in the variational circuits over all of the 20-node benchmark graphs. 
The structure of the 3-regular graphs is consistent across each random instance which leads to stable gate counts in the corresponding quantum circuits. The Erd\"os-R\'enyi graphs show much more variation and so the average number of each partial mixer within the circuits varies within and between each algorithm. Across all of the 20-node graphs, the DQVA circuits consistently contain the fewest number of partial mixers.\vspace{1em}

\twocolumn
\section{Gate Count Derivations for Decomposition Schemes}\label{app:asymptotic_counts}
In this section, we will provide references and derivations for each of the asymptotic gate count values provided in Table~\ref{tab:the-spreadsheet}.  We will break into cases depending on the number of ancillae being considered.

\subsection{No ancillae}
\begin{figure}[t]
\centering
\begin{quantikz}[column sep=0.33cm]
    \lstick[wires=5]{\(n-1\)} & \ctrl{6} & \ms{7} & \ctrl{5} & \qw            & \ctrl{5} & \qw      & \rstick[wires=3]{\(\ceil{\frac{n}{2}}\)}\qw \\
                              & \ctrl{5} & \qw    & \ctrl{4} & \qw            & \ctrl{4} & \qw      & \qw \\
                              & \ctrl{4} & \qw    & \ctrl{3} & \qw            & \ctrl{3} & \qw      & \qw \\
                              & \ctrl{3} & \qw    & \qw      & \ctrl{3}       & \qw      & \ctrl{3} & \rstick[wires=3]{\(\floor{\frac{n}{2}}\)}\qw \\
                              & \ctrl{2} & \qw    & \qw      & \ctrl{2}       & \qw      & \ctrl{2} & \qw \\
                              & \qw      & \qw    & \targ{}  & \ctrl{1}       & \targ{}  & \ctrl{1} & \qw \\
                              & \targ{}  & \qw    & \qw      & \targ{}         & \qw     & \targ{}  & \qw
\end{quantikz}
\caption{Splitting a multi-controlled not ``in half,'' into four multi-controlled not gates with half the number of controls \cite{gidney:2015:1}. Generalization of the decomposition scheme shown in Figure~\ref{fig:half}.}
\label{fig:half-cnot}
\end{figure}
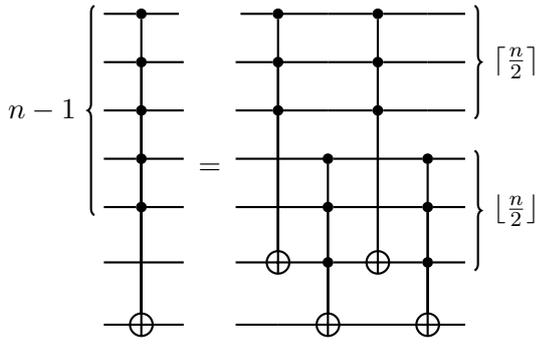

In the case where we have no ancillae, the gate counts for \(\mcS_2^2\) and \(\mcS_3^2\) come from Theorem 5 of \cite{iten2016quantum}. The asymptotic gate count for \(\mcS_m^2\) comes from first decomposing the \(C^n(R_x)\) gate into two \(C^{n-1}(X)\) gates via Lemma 7.9 of \cite{barenco1995elementary}. Then, we can split the \(C^{n-1}(X)\) gates in half using a similar circuit to that of Figure~\ref{fig:half}. The only difference is that the ancilla is now borrowed, so we need to split into four half-sized multi-controlled NOT gates, as shown in Figure~\ref{fig:half-cnot}.

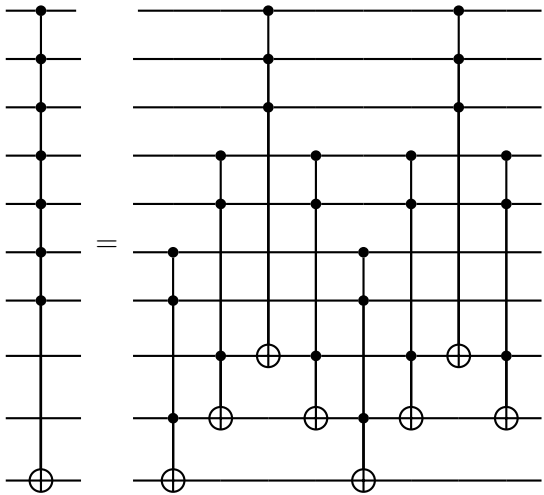
\begin{figure}[t]
    \centering
    \begin{quantikz}[column sep=0.3cm]
        & \ctrl{9} & \ms{10} & \qw      & \qw      & \ctrl{7} & \qw      & \qw      & \qw      & \ctrl{7} & \qw      & \qw \\
        & \ctrl{8} & \qw     & \qw      & \qw      & \ctrl{6} & \qw      & \qw      & \qw      & \ctrl{6} & \qw      & \qw \\
        & \ctrl{7} & \qw     & \qw      & \qw      & \ctrl{5} & \qw      & \qw      & \qw      & \ctrl{5} & \qw      & \qw \\
        & \ctrl{6} & \qw     & \qw      & \ctrl{5} & \qw      & \ctrl{5} & \qw      & \ctrl{5} & \qw      & \ctrl{5} & \qw \\
        & \ctrl{5} & \qw     & \qw      & \ctrl{4} & \qw      & \ctrl{4} & \qw      & \ctrl{4} & \qw      & \ctrl{4} & \qw \\
        & \ctrl{4} & \qw     & \ctrl{4} & \qw      & \qw      & \qw      & \ctrl{4} & \qw      & \qw      & \qw      & \qw \\
        & \ctrl{3} & \qw     & \ctrl{3} & \qw      & \qw      & \qw      & \ctrl{3} & \qw      & \qw      & \qw      & \qw \\
        & \qw      & \qw     & \qw      & \ctrl{1} & \targ{}  & \ctrl{1} & \qw      & \ctrl{1} & \targ{}  & \ctrl{1} & \qw \\
        & \qw      & \qw     & \ctrl{1} & \targ{}  & \qw      & \targ{}  & \ctrl{1} & \targ{}  & \qw      & \targ{}  & \qw \\
        & \targ{}  & \qw     & \targ{}  & \qw      & \qw      & \qw      & \targ{}  & \qw      & \qw      & \qw      & \qw
    \end{quantikz}
    \caption{Decomposition scheme for a multi-controlled \(X\) gate with access to two borrowed ancilla qubits that uses \(\mcnot{m-1}\) gates (here \(m-1 = 3\)) \cite{barenco1995elementary}. Generalization of the decomposition scheme shown in Figure~\ref{fig:cnot-into-toffoli} that uses gates larger than Toffoli's.}
    \label{fig:cnot-into-mcnot}
\end{figure}

Then, we can use a generalization of the decomposition scheme shown in Figure~\ref{fig:cnot-into-toffoli} where we decompose into \(\mcnot{m-1}\) gates instead of Toffolis. We show this scheme where \(m-1 = 3\) in Figure~\ref{fig:cnot-into-mcnot}. It is not too difficult to see that each \(\mcnot{k}\) gate requires \(\sim 4k/(m-2)\) \(\mcnot{m-1}\) gates to implement, and so putting this altogether the gate set \(\mcS_m^2\) uses
\begin{equation}
    4\qty(\frac{4}{m-2})\qty(\floor*{\frac{n}{2}} + \ceil*{\frac{n}{2}}) = {}\sim\!\qty(\frac{16}{m-2})n
\end{equation}
\(\mcnot{m-1}\) gates as desired.

\subsection{One ancilla}
In the case where we have one burnable ancilla and we want to decompose a \(C^n(R_x)\) gate, we can break up the gate into a \(C^{n-1}(X)\) gate and a \(C(R_x)\) gate to then decompose separately, as in Lemma 7.11 of \cite{barenco1995elementary} (here, since our ancilla is burnable, we don't need the extra \(C^{n-1}(X)\) gate). We can use the same decomposition scheme as shown in Figure~\ref{fig:half-cnot} to split the \(C^{n-1}(X)\) in half, and then use the decomposition scheme in Figure~\ref{fig:cnot-into-toffoli} to split it into Toffolis. The counts for \(\mcS_2^2\) and \(\mcS_3^2\) come from Lemma 8 of \cite{iten2016quantum} and Corollary 7.4 of \cite{barenco1995elementary}, respectively. The counts for \(\mcS_m^2\) can be computed in much the same way as we did for the no ancillae case -- since there is only one \(C^{n-1}(X)\) gate instead of two, we obtain the count of \(\sim (8/(m-2))n\) \(\mcnot{m-1}\) gates as desired.

In the case where we want to decompose a \(C^{n}(X)\) gate, we can just directly split the gate in half using the scheme in Figure~\ref{fig:half-cnot}, except we only need one of each ``half-sized'' multi-controlled not gate, since the ancilla is burnable. Then, the same procedures give the counts in the table for \(\mcS_2^2\) and \(\mcS_3^2\) as desired (notice also that now we do not have the extra \(C(R_x)\) gate to decompose, which is what gave the 4 extra single qubit gates and 2 extra two qubit gates in the \(\mcS_3^2\) case, by Corollary 5.3 of \cite{barenco1995elementary}). The counts for \(\mcS_m^2\) follow similarly, since we have again halved the number of \(C^{O(n)}(X)\) gates again.

\subsection{\(n\) ancillae}
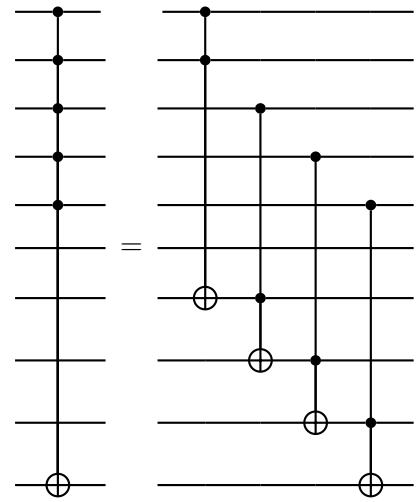
\begin{figure}[t]
    \centering
    \begin{quantikz}[column sep=0.4cm]
        & \ctrl{9} & \ms{10} & \ctrl{6} & \qw      & \qw      & \qw      & \qw \\
        & \ctrl{8} & \qw     & \ctrl{5} & \qw      & \qw      & \qw      & \qw \\
        & \ctrl{7} & \qw     & \qw      & \ctrl{5} & \qw      & \qw      & \qw \\
        & \ctrl{6} & \qw     & \qw      & \qw      & \ctrl{5} & \qw      & \qw \\
        & \ctrl{5} & \qw     & \qw      & \qw      & \qw      & \ctrl{5} & \qw \\
        & \qw      & \qw     & \qw      & \qw      & \qw      & \qw      & \qw \\
        & \qw      & \qw     & \targ{}  & \ctrl{1} & \qw      & \qw      & \qw \\
        & \qw      & \qw     & \qw      & \targ{}  & \ctrl{1} & \qw      & \qw \\
        & \qw      & \qw     & \qw      & \qw      & \targ{}  & \ctrl{1} & \qw \\
        & \targ{}  & \qw     & \qw      & \qw      & \qw      & \targ{}  & \qw
    \end{quantikz}
    \caption{Modified decomposition scheme from Figures \ref{fig:cnot-into-toffoli} and \ref{fig:cnot-into-mcnot} that now uses burnable ancillae. As a result, we don't have to return the ancillae to their original states, allowing us to quarter the number of gates required.}
    \label{fig:cnot-burnable}
\end{figure}

In the case where we have \(n\) burnable ancillae, we can actually modify the decomposition scheme in Figures~\ref{fig:cnot-into-toffoli}~and~\ref{fig:cnot-into-mcnot} to use about one fourth as many Toffolis. We show this modified scheme in Figure~\ref{fig:cnot-burnable}. This gives us the \(n-1\) count for the gate set \(\mcS_3^2\), but for the other counts, we need to treat the last Toffoli as a \(C^2(R_x)\) gate. The same techniques as those above give us the remaining counts. The gate counts for \(\mcS_m^2\) come from generalizing the decomposition scheme from Figure~\ref{fig:cnot-burnable} in the same way as we did to get Figure~\ref{fig:cnot-into-mcnot}.

\subsection{Gate counts for \(\mcS_2^3\)}
The general tree-like decomposition for a multi-controlled unitary using qutrits is given in \cite{gokhale2019asymptotic} (see Figure 5 of \cite{gokhale2019asymptotic}).
The three-qutrit gates that appear in the decomposition --- consisting of two controls on either the $\ket{1}$ or $\ket{2}$ qutrit states and an increment or decrement operation applied to the final qutrit --- are then further decomposed into single- and two-qutrit operations using the decomposition found in Figure 10 of \cite{di2011elementary}.
For each three-qutrit operation, the decomposition of \cite{di2011elementary} yields 6 two-qutrit and 7 single-qutrit operations. Therefore, for a multi-controlled gate with $n$ controls, $n-1$ three-qutrit operations will appear due to the decomposition of \cite{gokhale2019asymptotic}, and each of these yields 6 two-qutrit operations due to \cite{di2011elementary}, finally two extra two-qutrit operations are required to implement the single-controlled unitary. This yields results in an asymptotic scaling of $6n - 4$ two-qutrit operations required to implement $n$-controlled $R_x$ gates. The single-qutrit gate costs similarly follow using the same decomposition given in \cite{di2011elementary}.

\end{document}